\documentclass{jfm}
\usepackage{graphicx}
\usepackage{dcolumn}
\usepackage{bm}
\usepackage{color}
\usepackage{placeins}
\usepackage{array}
\usepackage{subfigure}
\usepackage{natbib}
\usepackage{amssymb}
\usepackage{amsbsy}
\usepackage{graphicx}
\usepackage{amssymb}
\usepackage{amsmath}

\begin{document}

\title[LES and log-law for second and higher-order moments in wall-bounded turbulence]
{Large-eddy simulation study of the logarithmic law for second and higher-order moments in turbulent wall-bounded flow}

\author[Richard J.A.M. Stevens, Michael Wilczek and Charles Meneveau]
{Richard J.A.M. Stevens$^{1,2}$, Michael Wilczek$^1$ and Charles Meneveau$^1$}

\affiliation{
$^1$Department of Mechanical Engineering, Johns Hopkins University, Baltimore, Maryland 21218, USA.\\
$^2$Department of Science and Technology and J.M. Burgers Center for Fluid Dynamics, University of Twente, P.O Box 217, 7500 AE Enschede, The Netherlands.}

\pubyear{}
\volume{}
\pagerange{}
\date{\today}

\maketitle

\begin{abstract}
The logarithmic law for the mean velocity in turbulent boundary layers has long provided a valuable and robust reference for comparison with theories, models, and large-eddy simulations (LES) of wall-bounded turbulence. More recently, analysis of high-Reynolds number experimental boundary layer data has shown that also the variance and higher-order moments of the streamwise velocity fluctuations $u'^{+}$ display logarithmic laws. Such experimental observations motivate the question whether LES can accurately reproduce the variance and the higher-order moments, in particular their logarithmic dependency on distance to the wall. In this study we perform LES of very high Reynolds number wall-modeled channel flow and focus on profiles of variance and higher-order moments of the streamwise velocity fluctuations. In agreement with the experimental data, we observe an approximately logarithmic law for the variance in the LES, with a `Townsend-Perry' constant of $A_1\approx 1.25$. The LES also yields approximate logarithmic laws for the higher-order moments of the streamwise velocity. Good agreement is found between $A_p$, the generalized `Townsend-Perry' constants for moments of order $2p$, from experiments and simulations. Both are indicative of sub-Gaussian behavior of the streamwise velocity fluctuations. The near-wall behavior of the variance, the ranges of validity of the logarithmic law and in particular possible dependencies on characteristic length scales such as the roughness scale $z_0$, the LES grid scale $\Delta$, and sub-grid scale (SGS) mixing length $C_s\Delta$ are examined. We also present LES results on moments of spanwise and wall-normal fluctuations of velocity. 
\end{abstract}

\section{Introduction}
The logarithmic law of the wall for the mean velocity in a rough-wall turbulent boundary layer, written below using an effective roughness scale,
\begin{equation} \label{equation1}
 \frac{\langle u \rangle}{u_*} = \frac{1}{\kappa} \log\left({\frac{z}{z_0}}\right),
\end{equation}
is a well-established result \citep{pra25,kar30,mil38}. Here, $u$ is the streamwise velocity component, $u_*$ is the friction velocity, $\kappa \approx 0.4$ is the von K\'arm\'an constant, $z$ is the height from the wall, and $z_0$ is the roughness length.

Models based on the `attached-eddy hypothesis' \citep{tow76,per82,per86} have predicted a logarithmic behavior for the variance of the fluctuations of the streamwise velocity component in the inertial layer. However, only recently clear experimental evidence \citep{mar03,hul12,mar13} has emerged for a universal law for the variance (second-order moment profiles) of the streamwise velocity fluctuations, based on well-resolved experimental boundary layer data at sufficiently high Reynolds numbers. The log-law for the variance has the form
\begin{equation} \label{equation2}
\langle (u'^{+})^2 \rangle = B_1 -A_1 \log\left({\frac{z}{\delta}}\right),
\end{equation}
where $u'^{+}$=$(u - \langle u \rangle)/u_*$ is the normalized streamwise velocity fluctuation and $\delta$ is an outer length scale. The experimental data are consistent with a value of $A_1 \approx 1.25$, i.e. the `Townsend-Perry' constant \citep{mar03,smi11,hul12,mar13,men13}, while $B_1$ depends on the flow conditions and geometry and is not thought to be universal. When Gaussian behavior is assumed, the even-order moments can be related to the second-order moment through the relationship $\langle (u'^{+})^{2p} \rangle = (2p-1)!! \langle (u'^{+})^{2} \rangle^p$, where $n!! \equiv n (n-2) (n-4) \dots 2$ is the double factorial \citep{men13}. This relationship between the second and $2p^\mathrm{th}$-moments means that the $p^\mathrm{th}$ root of the even order moments of the velocity fluctuations should follow a generalized logarithmic law for higher-order moments as follows
\begin{equation} \label{equation3}
\langle (u'^{+})^{2p}\rangle^{1/p} = B_p -A_p \log\left({\frac{z}{\delta}}\right).
\end{equation}
The assumption of Gaussian statistics furthermore implies that $A_p=A_1 [(2p-1)!! ]^{1/p}$. The results of \cite{men13} show that the experimental data are consistent with the logarithmic trends of the $p$th root of the moments, but deviations from the Gaussian prediction for the slopes $A_p$ are found. 

Observation of such possibly canonical statistical behavior in boundary layers provides valuable points of reference for turbulence theories and various applications. Knowledge about the probability density  function of velocity fluctuations plays an important role in diverse practical applications, such as characterizing wind-turbine power fluctuations to estimating probabilities of extreme events. In addition, the generalized logarithmic laws for higher-order moments may serve as a new benchmark on which to test predictions from models and simulations. 

There is relatively little information available about the ability of LES to reproduce accurately higher-order statistics of turbulence. Most of the literature to date focuses on comparisons of mean velocity distributions and second-order moments. It is important to recall that most LES models are motivated by the need to dissipate kinetic energy at the correct rate, i.e. to reproduce the correct second-order statistics such as mean kinetic energy. However, there is no guarantee that the inherent nonlinear dynamics of LES will actually reproduce the extreme values of the distributions that arise from the real nonlinear dynamics in the real physical system. An earlier study \citep{kan03} provided comparisons of LES and experiments for inertial-range velocity increments and their high-order moments in decaying isotropic turbulence. Overall, the results were encouraging. However, in wall-bounded flows the situation is significantly more challenging due to flow inhomogeneity, anisotropy, wall-blocking, etc. Prior studies on the accuracy of LES for high-Reynolds number wall-modeled turbulent boundary layers include those of \cite{bra10}, who explored various resolution criteria to reproduce accurately the mean velocity profiles, and \cite{sul11} who documented behavior of variances and third-order moments. In this paper, we use data from high-resolution LES of a turbulent wall-bounded flow to study the ability of LES to reproduce fundamental scaling laws for second as well as higher-order moments. Such analysis has not yet been done and is needed to place LES on firmer fundamental ground as a tool to model turbulence. 

In section \ref{section2} we start with a brief description of the simulation method. Subsequently, in section \ref{section3_1} we compare the streamwise velocity fluctuation variance from LES with experimental data \citep{hut09,men13} and in section \ref{section3_2} we discuss the role of the numerical resolution and possible effects of model and physical length scales characterizing the near-wall region and in setting the lower limit of the logarithmic region for the variance. Then, in section $\ref{section3_4}$ the spanwise and vertical velocity fluctuations are analyzed in more detail, which is followed by conclusions in section \ref{section4}.

\begin{figure}
\centering
\subfigure{\includegraphics[width=0.99\textwidth]{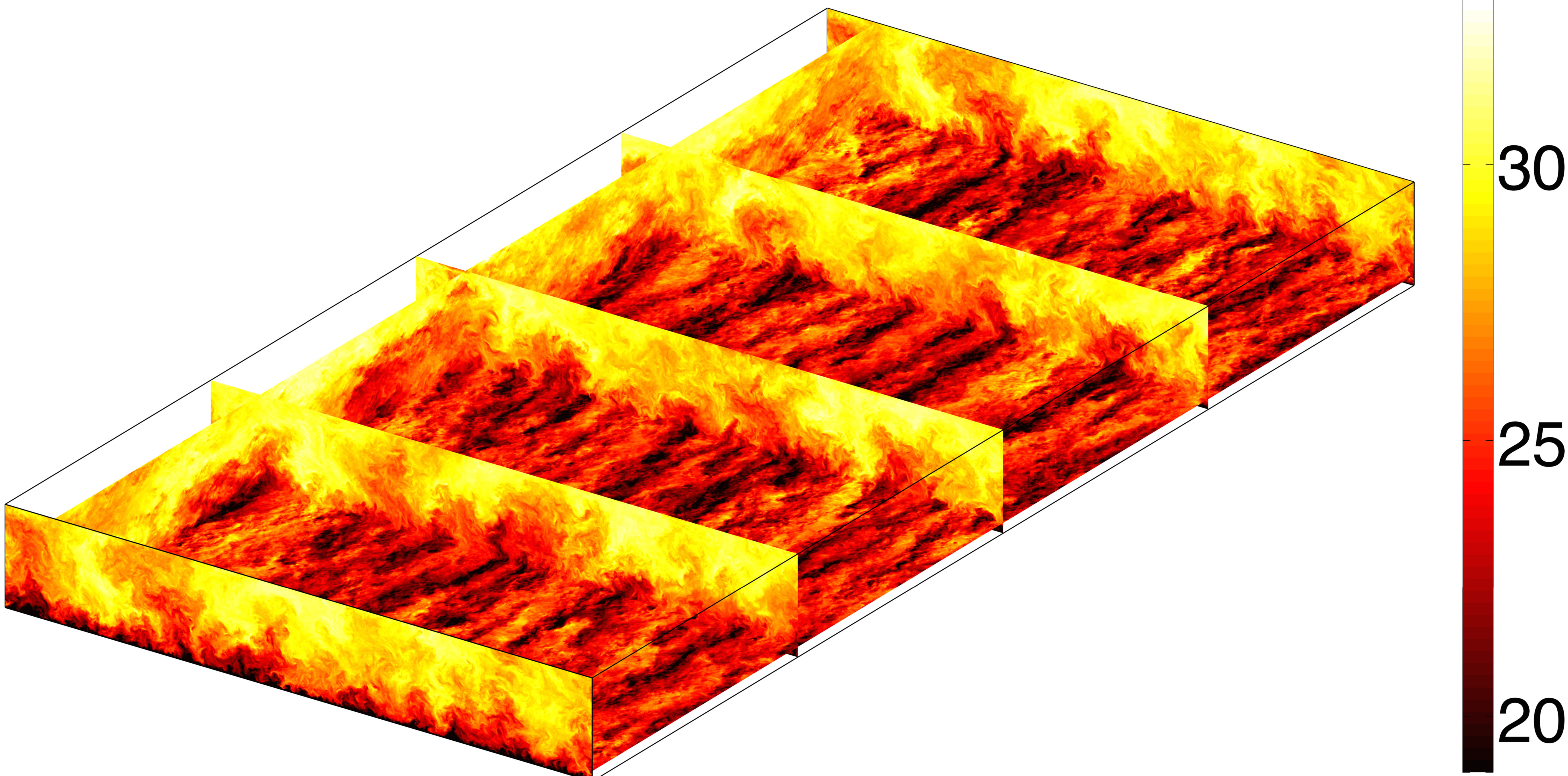}}
\caption{Snapshot of the streamwise velocity from the LES performed on a $2048 \times 1024 \times 577$ grid with $z_0/H=1\times10^{-5}$ (Case I2). The color indicates the streamwise velocity in non-dimensional units $u/u_*$.}
\label{figure1}
\end{figure}

\section{Large-eddy simulations} 
\label{section2}
The LES code we use to study the turbulent wall-bounded flow solves the filtered incompressible Navier-Stokes equations without buoyancy, system rotation or other effects. The nonlinear terms are evaluated in rotational form. A pseudo-spectral discretization and thus double periodic boundary conditions are used in the horizontal directions parallel to the wall, while centered second-order finite differencing is used in the vertical direction \citep{moe84,alb99,por00}. The deviatoric part of the sub-grid scale stress term is modeled using an eddy-viscosity sub-grid scale model, employing the scale-dependent Lagrangian dynamic approach in conjunction with the Smagorinsky model and a sharp spectral cutoff test-filter \citep{bou05}. Only this model will be used here, since this study is not focused on comparing the performance of different sub-grid scale models (such comparisons will be presented elsewhere). The trace of the SGS stress is combined into the modified pressure, as is common practice in LES of incompressible flow. A second-order accurate Adams-Bashforth scheme is used for the time integration. Due to the very large Reynolds numbers considered here we parameterize the bottom surface by using a classic imposed wall stress boundary condition. This boundary condition relates the wall stress to the velocity at the first grid point using the standard logarithmic similarity law \citep{moe84} using velocities test-filtered at twice the grid scale \citep{bou05}. This test-filtering ensures that the average predicted stress is close to the stress predicted by the classic logarithmic law. In addition the viscous stresses are neglected. 

The wall stress is expressed in terms of the velocity at the first grid point above the wall (at height $\Delta z/2$ for the staggered vertical mesh, $\Delta z$ is the vertical grid spacing) according to 
\begin{equation} \label{equation4}
\tau_w(x,y)= - \left[ \frac{\kappa}{\log \left[ (\Delta z/2)/z_0 \right] } \right]^2 \left( \left[\overline{u}(x,y,\Delta z/2) \right]^2 +\left[\overline{v}(x,y,\Delta z/2) \right]^2 \right),
\end{equation}
where $\overline{u}$ and $\overline{v}$ indicate the test-filtered, with a spectral cutoff, streamwise and spanwise velocity. 
Subsequently, the stress is divided into its streamwise and spanwise component using the direction of $\overline{\bf u}(x,y,\Delta z/2)$. For the top boundary we use a zero-vertical-velocity and zero-shear-stress boundary condition so that the flow studied corresponds effectively to a `half-channel flow' with an impermeable centerline boundary. The flow is driven by an applied pressure gradient in the $x$-direction, which in equilibrium determines the wall stress $u_*^2$ and the velocity scale $u_*$ used to normalize the results of the simulations, together with the domain height $H$ used to normalize length scales. The LES code has been further documented and applied in various previous publications \citep{por00,bou05,che07,cal10,cha11}. 
\begin{figure}
\centering
\subfigure[]{\includegraphics[width=0.49\textwidth]{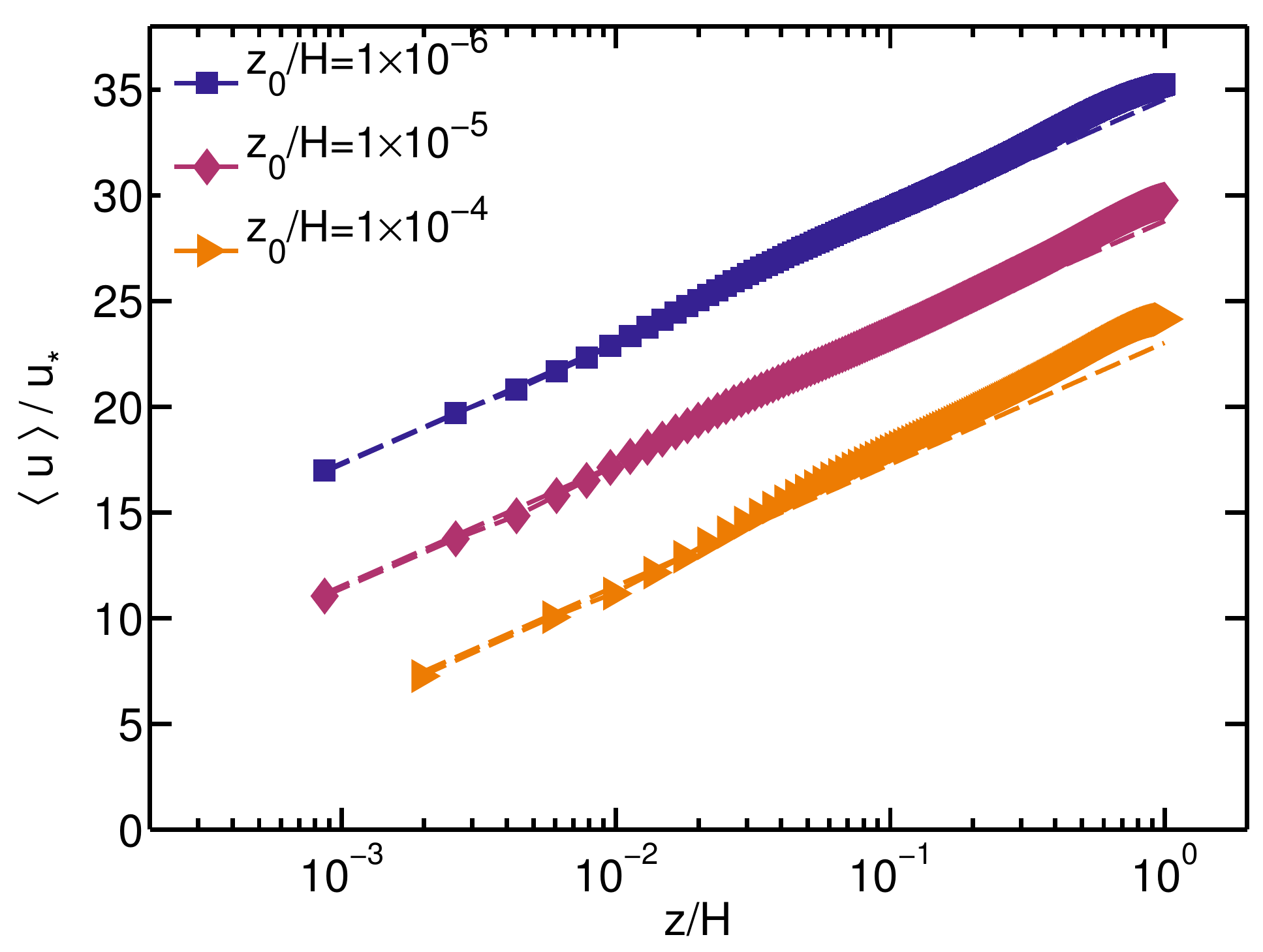}}
\subfigure[]{\includegraphics[width=0.49\textwidth]{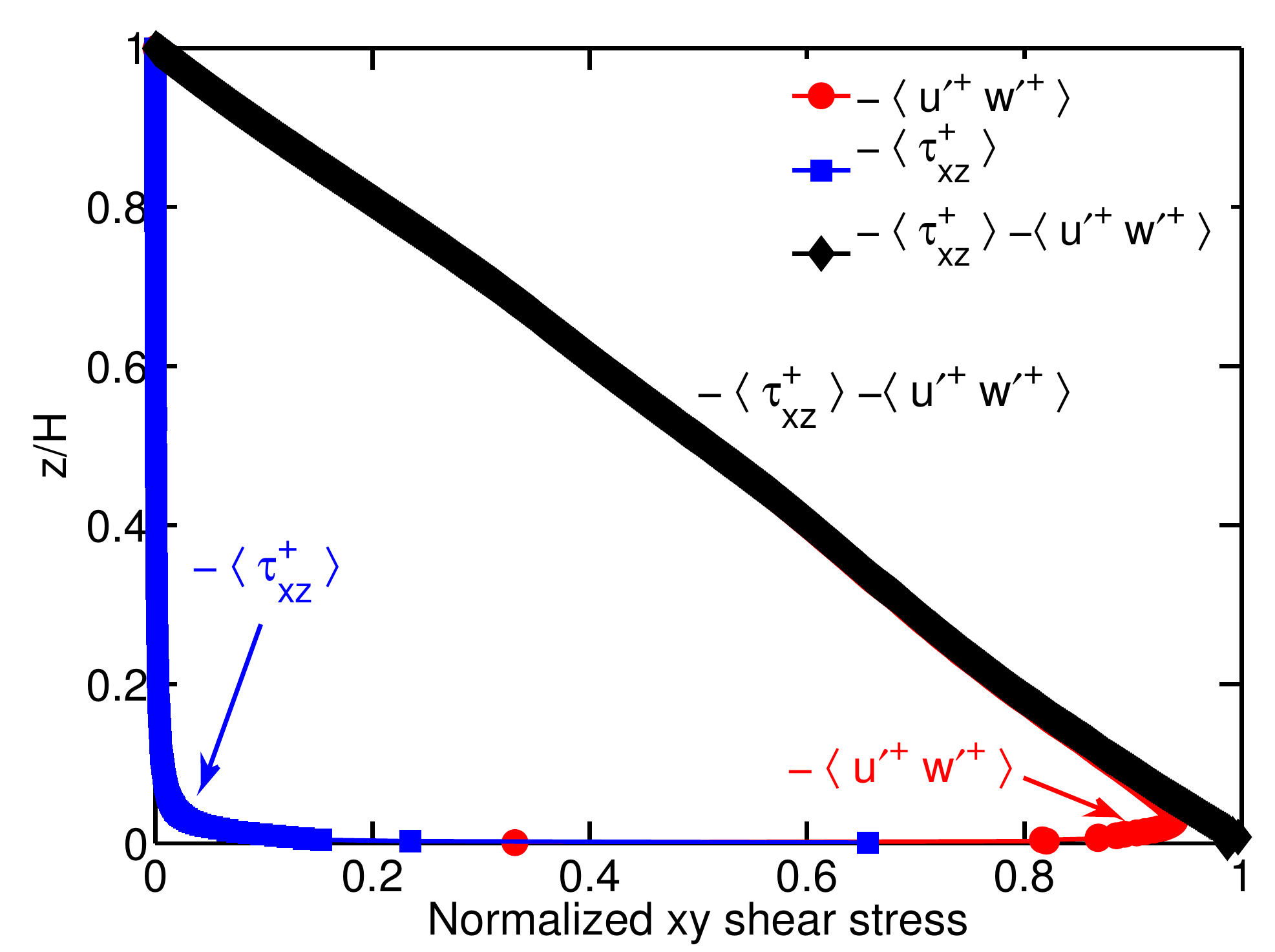}}
\caption{(a) Horizontally averaged streamwise velocity from LES compared with the logarithmic law for the mean flow. The squares, diamonds, and triangles indicate the results for $z_0/H=1\times10^{-6}$, $z_0/H=1\times10^{-5}$, and $z_0/H=1\times10^{-4}$, respectively, while the dashed lines are the expected corresponding log-laws. (b) Vertical profiles of the resolved stress ($- \langle u'^{+}w'^{+} \rangle$, circles) the normalized sub-grid scale stress ($-\langle \tau_{xz}^+ \rangle$, squares) and the total stress ($-\langle u'^{+}w'^{+} \rangle-\langle \tau_{xz}^+ \rangle$, diamonds) for $z_0/H=1\times10^{-5}$, for case I2 (on a $2048 \times 1024 \times 577$ grid).}
\label{figure2}
\end{figure}

\begin{table}
\centering
\caption{Large-eddy simulation cases. The simulation for $z_0/H=10^{-4}$ and $z_0/H=10^{-5}$ are performed on a $(4\pi \times 2 \pi \times 1)H$ domain in the streamwise, spanwise and vertical direction, respectively, and the $z_0/H=10^{-6}$ cases on a $(6\pi \times 3 \pi \times 1)H$ domain. The table indicates the number of grid points used for the different cases and the ratio between the horizontal and vertical grid spacing. All simulations use the scale-dependent Lagrangian sub-grid model. }
\FloatBarrier
 \begin{tabular}{|c|c|c|c||c|c|c|c|c|}
 \hline
 name 	& $z_0/H$ 	& 	$N_x \times N_y \times N_z$		&	$\frac{\Delta x=\Delta y}{\Delta z}$ 	& name 	& $z_0/H$ 		& 	$N_x \times N_y \times N_z$		&	$\frac{\Delta x=\Delta y}{\Delta z}$\\	
 \hline 
 A1		& $10^{-4}$	& 	$64 \times 32 \times 32$ 			&	$2.00\pi$						&	 E2		& $10^{-5}$	& 	$128 \times 64 \times 32$ 		&	$1.00\pi$	\\	
 A2		& $10^{-4}$	& 	$128 \times 64 \times 32$ 		&	$1.00\pi$						&	 F2		& $10^{-5}$	&	$256 \times 128 \times 64$		&	$1.00\pi$	\\	
 B1		& $10^{-4}$	&	$128 \times 64 \times 64$ 		&	$2.00\pi$						&	 G2		& $10^{-5}$	&	$512 \times 256 \times 128$		& 	$1.00\pi$	\\
 B2		& $10^{-4}$	&	$256 \times 128 \times 64$		&	$1.00\pi$						&	 H2		& $10^{-5}$	&	$1024 \times 512 \times 256$		& 	$1.00\pi$	\\
 B3		& $10^{-4}$	&	$512 \times 256 \times 64$ 		&	$0.50\pi$						&	 I2		&$10^{-5}$	&	$2048 \times 1024 \times 577$		& 	$1.13\pi$	\\	
 C1		& $10^{-4}$	&	$256 \times 128 \times 128$		& 	$2.00\pi$						&	 J2		& $10^{-6}$	& 	$192 \times 96 \times 32$ 		&	$1.00\pi$	\\		
 C2		& $10^{-4}$	&	$512 \times 256 \times 128$		& 	$1.00\pi$						&	 K2		& $10^{-6}$	&	$384 \times 192 \times 64$		&	$1.00\pi$	\\	 
 C3		& $10^{-4}$	&	$1024 \times 512 \times 128$		& 	$0.50\pi$						&	 L2		& $10^{-6}$	&	$768 \times 384 \times 128$		& 	$1.00\pi$	\\
 D1		& $10^{-4}$	&	$512 \times 256 \times 256$		& 	$2.00\pi$						&	 M2		& $10^{-6}$	&	$1536 \times 768 \times 256$		& 	$1.00\pi$	\\
 D2		& $10^{-4}$	&	$1024 \times 512 \times 256$		& 	$1.00\pi$						& 	 N2		& $10^{-6}$	&	$2048 \times 1024 \times 577$		& 	$1.69\pi$	\\	
 \hline
 \end{tabular} 
\label{table1}
\end{table}

As periodic boundary conditions are used in the streamwise and spanwise directions, a sufficiently large domain in these directions is required in order to allow the flow to develop with negligible correlation over the domain length. Therefore we use a domain up to $(6\pi \times 3 \pi \times 1)H$ in the streamwise, spanwise, and vertical directions, respectively. We perform simulations with different grid resolutions and roughness scale $z_0/H$, see table \ref{table1}, to study their influence. Note that for the $z_0/H=10^{-6}$ case a larger domain size is used. For this case the mean velocities (compared to the friction velocity) are higher than for the higher roughnesses and, as will be explained in more detail below, we found that a larger computational domain is necessary for this case to prevent unphysical streamwise and spanwise correlations associated to the use of periodic boundary conditions. As discussed below in detail, a sufficient grid resolution is needed to accurately capture the logarithmic region for higher-order moments. In the beginning of the article we compare the simulation performed on the $2048 \times 1024 \times 577$ grid with $z_0/H=10^{-5}$ (case I2) with the smooth-wall experimental data collected at the University of Melbourne (hereafter the Melbourne data) at $Re_\tau=19,030$ \citep{hut09} before we compare the LES of the cases with different roughness lengths with several experimental data sets.

In order to limit the computational time that is needed to reach a statistically stationary state, an interpolated flow field obtained from a lower resolution simulation is used as initial condition for the next, higher resolution simulation. Each case has been run for about $100$ dimensionless time units (where the dimensionless time is in units of $H/u_*$) before it is used as initial condition. For the simulation cases on the $1024\times512\times256$ grid this is followed by integrating for an additional $1$ dimensionless time unit on the fine grid before data collection is started. Subsequently, data are collected over roughly $2.5$ time units while collecting a full snapshot of the flow field every $\approx 0.07$ time units. For the $2048\times1024\times577$ cases the flow is simulated for approximately $1$ time unit and snapshots are saved every $0.03$ time units.

A visualization of streamwise velocity normalized by $u_*$ from a snapshot is shown in figure \ref{figure1}. The usual elongated structures can be seen at various distances to the wall and the increase of the variance towards the wall is also evident. In figure \ref{figure2}a we show that the time averaged mean velocity is close to the expected logarithmic law for the three different roughness lengths considered here, although a small ``bump'' in the log-law can be discerned at about $z/H \approx 0.02$ to $0.03$, depending on the grid resolution. Many works have proposed various improvements in sub-grid and wall modeling approaches to further improve agreement with the logarithmic law for mean velocity. Here we take a well-documented model which exhibits good (but not optimized) performance in predicting the mean velocity and focus on higher-order moments of the fluctuating (resolved) velocity. The horizontally averaged vertical stress profiles in figure \ref{figure2}b confirm that the flow has reached a statistically stationary state in the sense that the linear shear stress profile balances the driving pressure gradient. Figure \ref{figure2}b shows that, due to the high resolution, the modeled normalized sub-grid scale stresses ($-\langle \tau_{xz}^+ \rangle$) only become larger than $10\%$ of the total stresses ($- \langle \tau_{xz}^+ \rangle - \langle u'^{+}w'^{+} \rangle$) for $z/H \lesssim 0.01$, for the case shown (case I2). This is also the location where the ``bump'' is seen in the mean velocity profile. We note that the results for the other roughness lengths are similar.

\begin{figure}
\centering 
\subfigure[]{\includegraphics[width=0.49\textwidth]{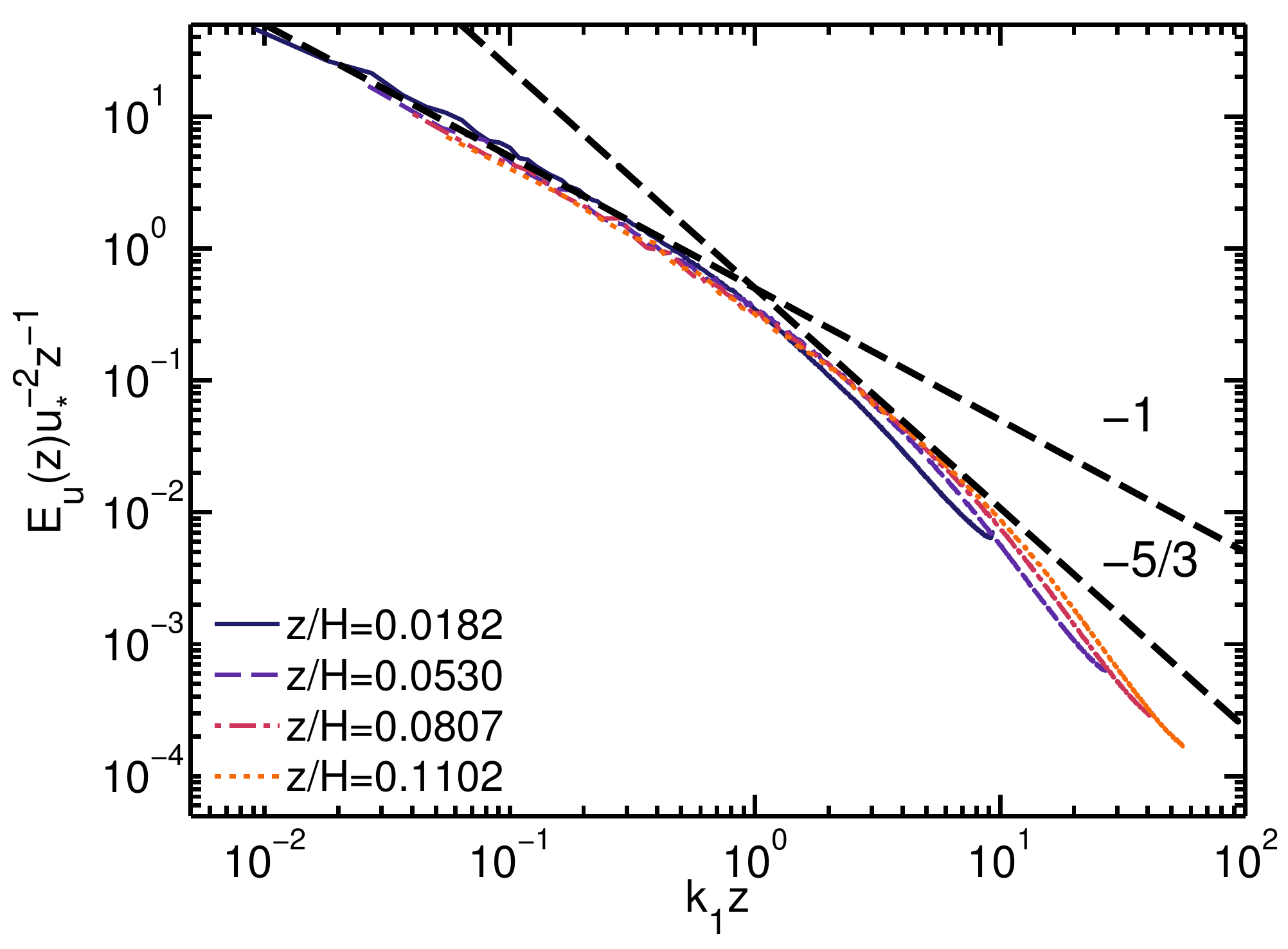}}
\subfigure[]{\includegraphics[width=0.49\textwidth]{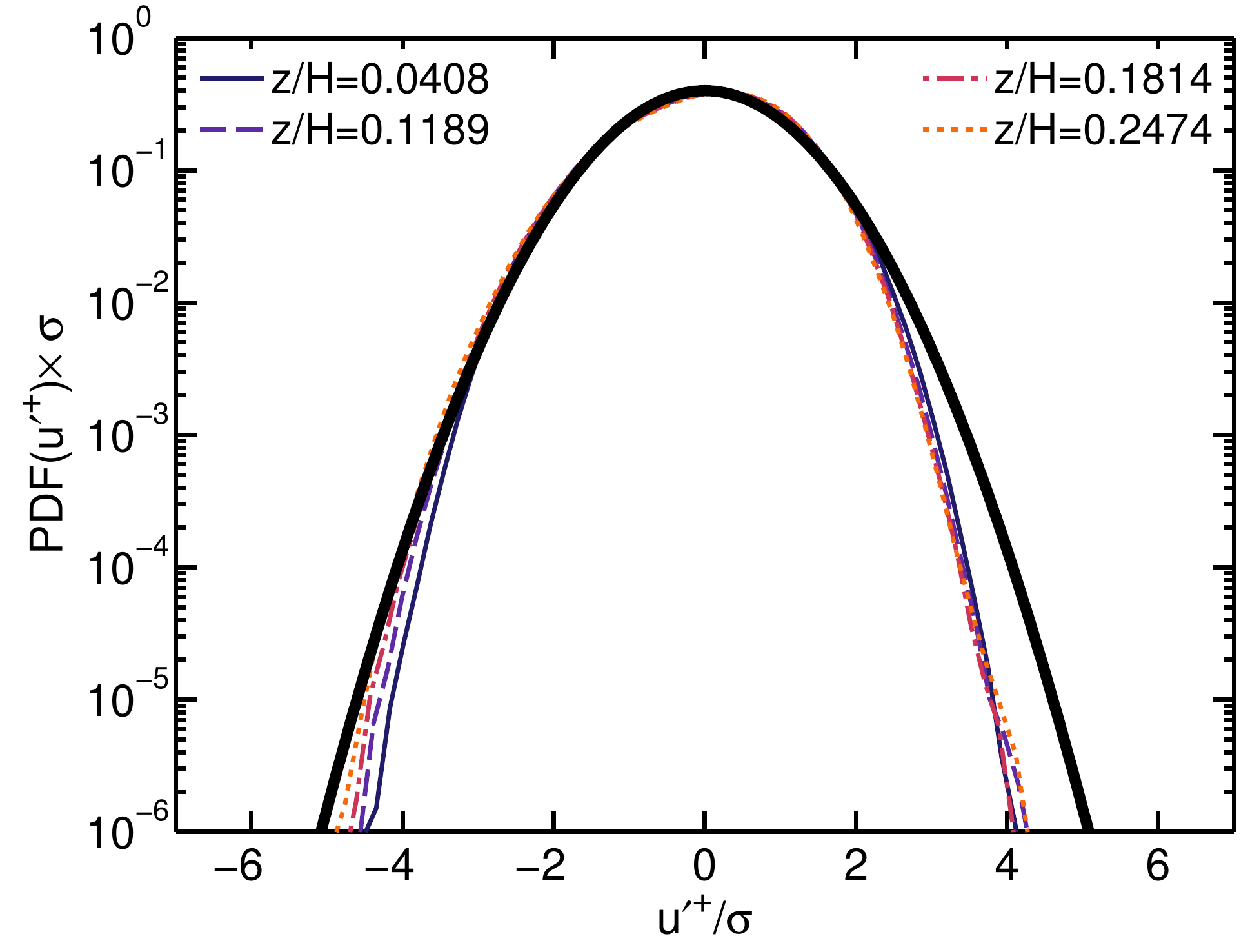}}
\subfigure[]{\includegraphics[width=0.49\textwidth]{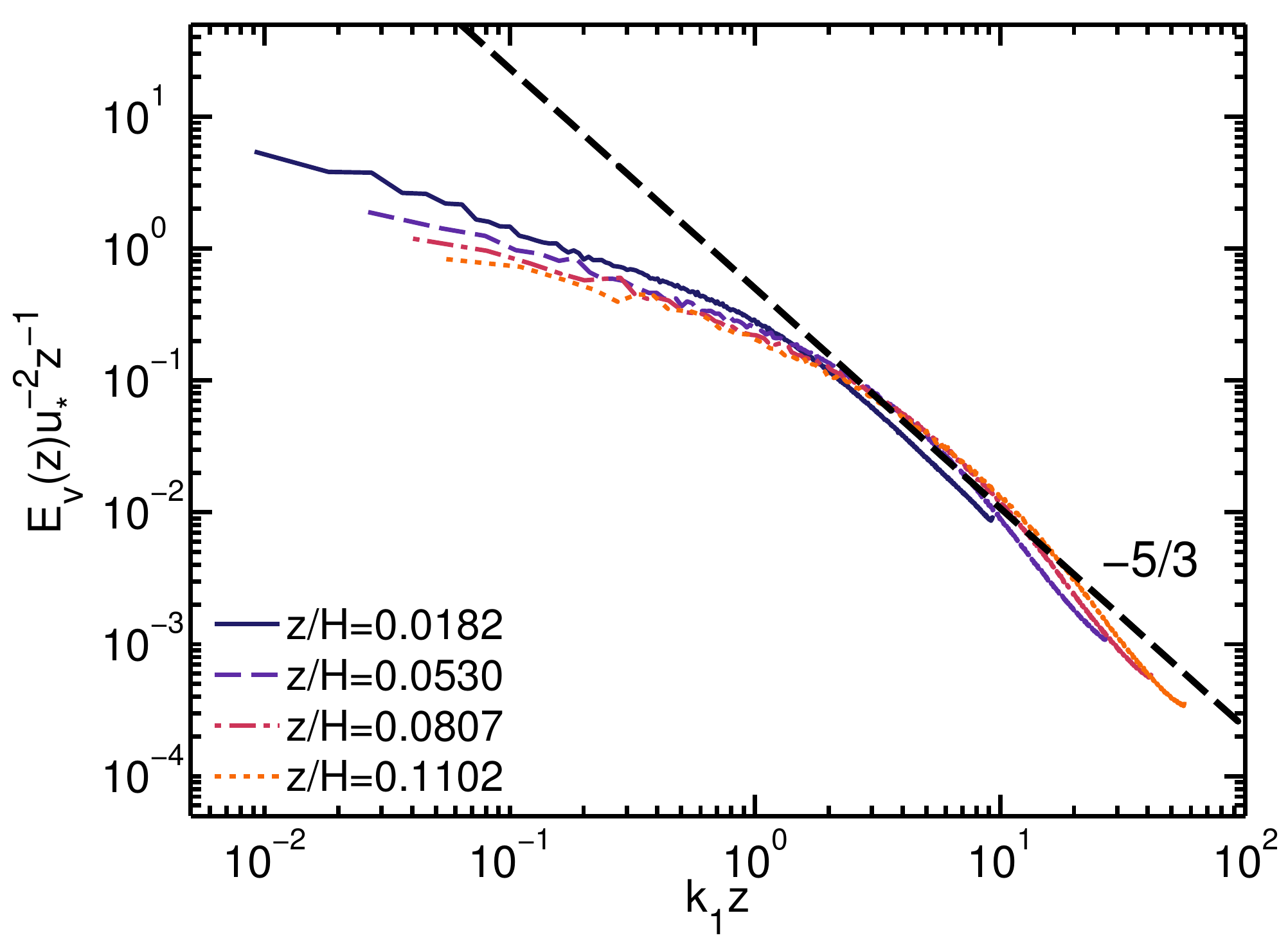}}
\subfigure[]{\includegraphics[width=0.49\textwidth]{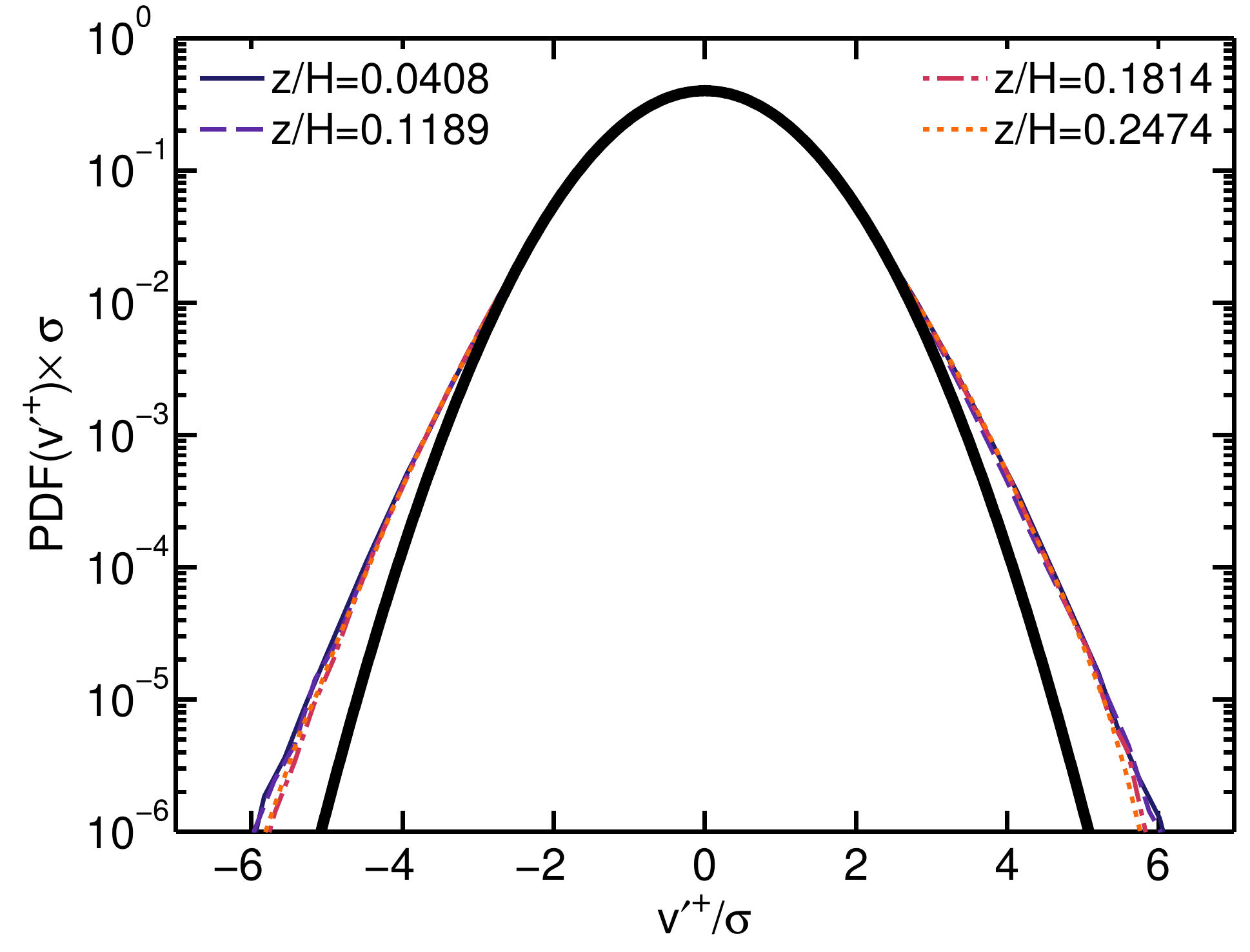}}
\subfigure[]{\includegraphics[width=0.49\textwidth]{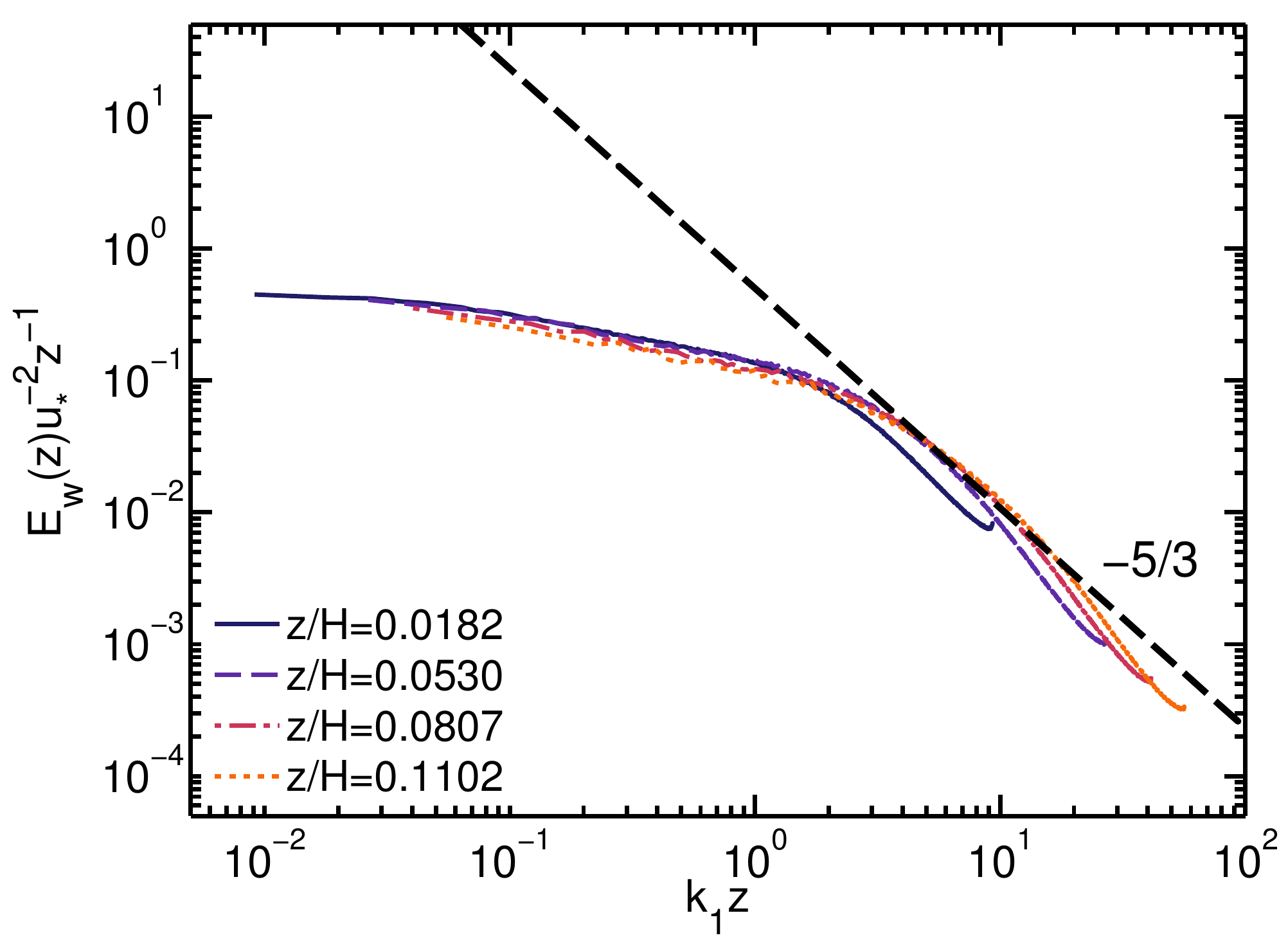}}
\subfigure[]{\includegraphics[width=0.49\textwidth]{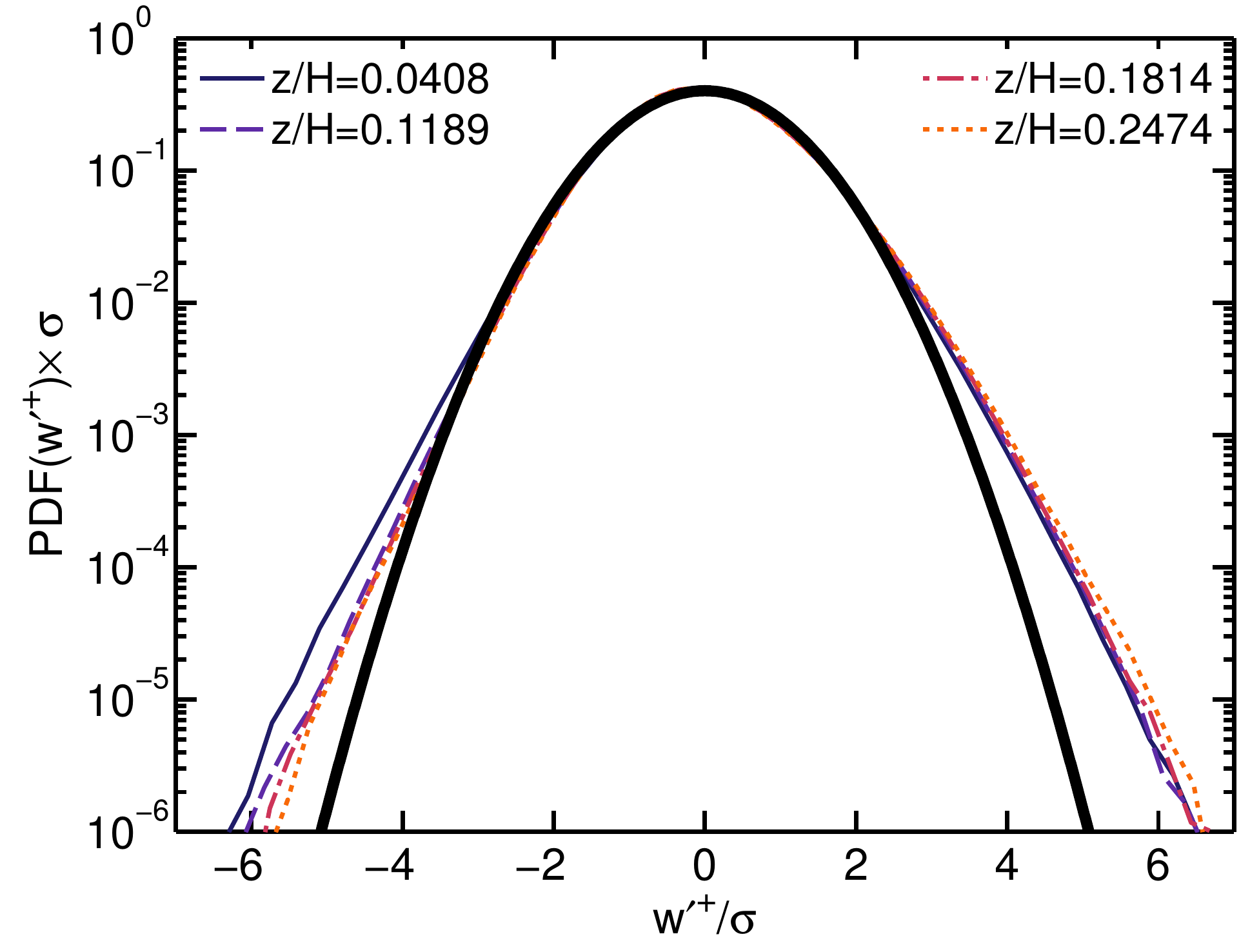}}
\caption{Streamwise spectra of the (a) streamwise velocity (c) spanwise velocity, and (e) vertical velocity component. Also shown are probability density functions (PDF's) for the (b) streamwise, (d) spanwise and (e) vertical velocity fluctuations at different heights. The dark line is a Gaussian distribution shown as reference.}
\label{figure3}
\end{figure}

Further characterization of the LES result for case I2 is provided from the streamwise spectra shown in figure \ref{figure3}. The LES captures some of the expected $k^{-5/3}$ behavior for all velocity components, although the spectra become slightly steeper for the highest wave-numbers. A peel-off at various heights can be seen that prevents a complete collapse onto a single -5/3 slope due to the different values of the cutoff wavenumber when normalized with $z$. In the production range $k_1z<1$, for the streamwise velocity component a slope of approximately $-1$ is observed in experiments \citep{per87} and in LES of atmospheric boundary layers by \cite{por00}, see figure \ref{figure3}a. The slope is lower for the spanwise velocity component and approximately horizontal for the vertical velocity component.

\begin{figure}
\centering
\subfigure[]{\includegraphics[width=0.49\textwidth]{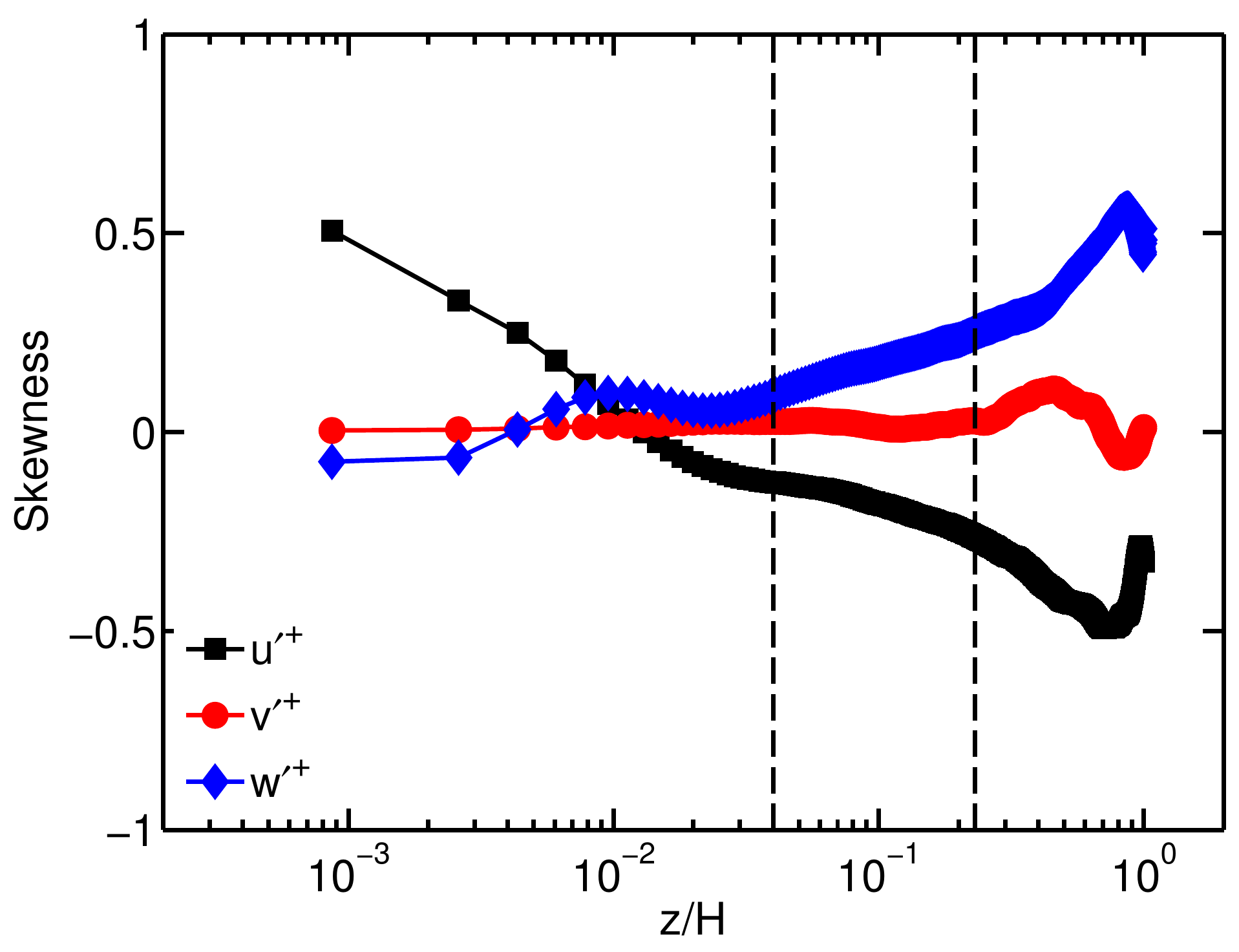}}
\subfigure[]{\includegraphics[width=0.49\textwidth]{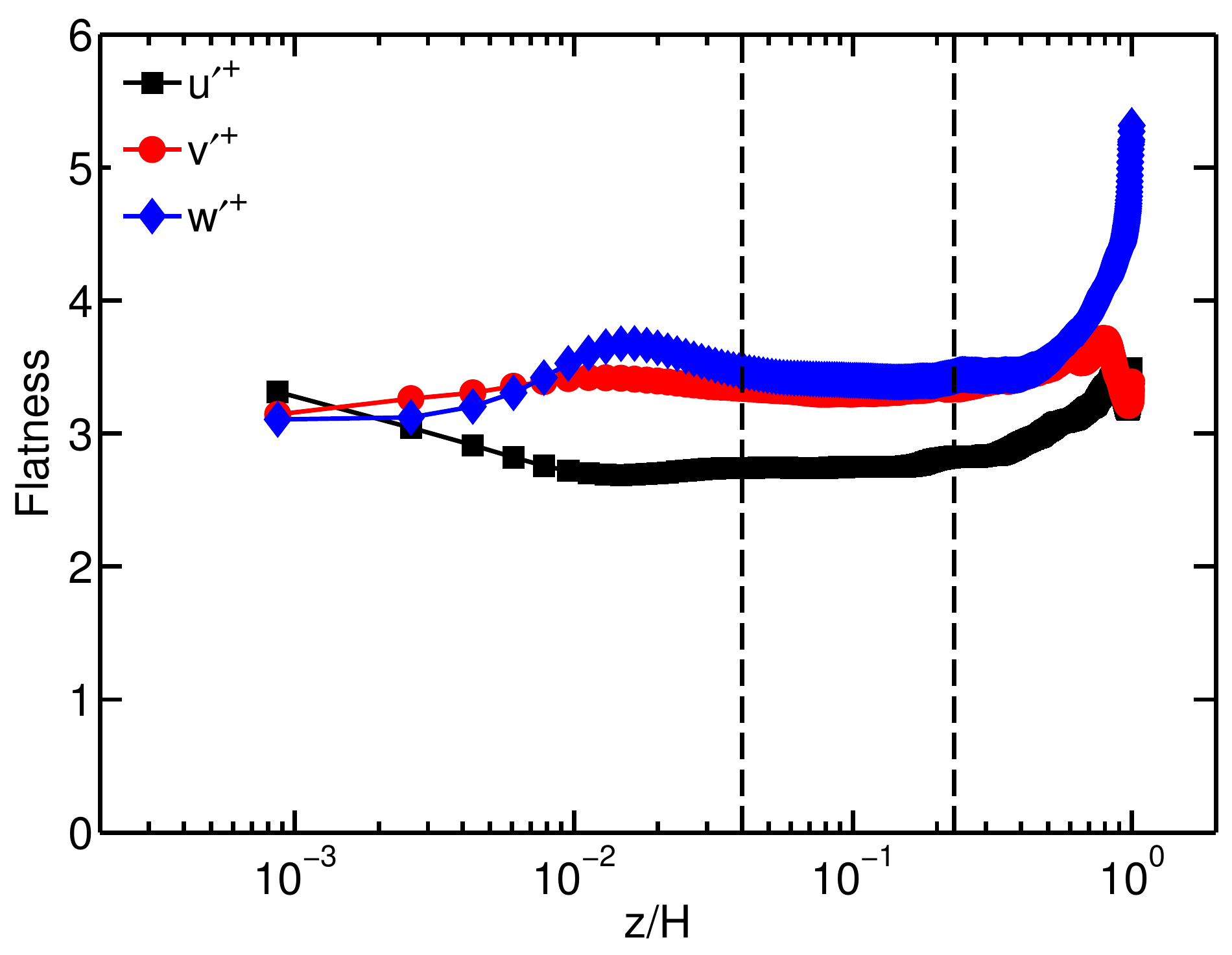}}
\caption{(a) Skewness and (b) flatness of the streamwise ($u'^{+}$), spanwise ($v'^{+}$) and vertical ($w'^{+}$) velocity component as function of $z/H$. The dashed vertical lines indicate the region ($0.04 \leq z/H \leq 0.23$).}
\label{figure4}
\end{figure}

The PDFs of the velocity fluctuations in figure \ref{figure3} reveal that the streamwise velocity component is sub-Gaussian, hence with a negative skewness, see figure \ref{figure4}a. The skewness for the three velocity components is plotted in figure \ref{figure4}a. The sign change of the skewness for the streamwise velocity component has also been observed in experimental data \citep{met01,mar10} and a corresponding model has been proposed by \cite{mar10}. The spanwise and vertical velocity components are super-Gaussian, with a positive skewness for the vertical velocity component and a near zero skewness for the spanwise velocity component. For the vertical velocity component we can compare the results observed in LES of planetary boundary layers, see e.g.\ \cite{sul11} and \cite{moe90}, which is a slightly different case because the effect of the temperature inversion is not included in our simulations. \cite{sul11} show that the change from negative to positive skewness shifts from $z/z_i \approx 0.1$ on a $32^3$ grid, where $z_i$ is the inversion height, towards $z/z_i \lesssim 0.01$ on a $1024^3$ grid. The same trend is observed in our dataset. Apart from this near wall behavior, the LES results and measurements \citep{moe90,len12} are found to be consistent although we note that there is significant scatter in the measurements. The observed flatness of the vertical velocity component between $3$ and $4$ for $0.05 \lesssim z/H \lesssim 0.5$ and its increase at the top of the domain are also consistent with the LES of \cite{sul11} and recent measurements \citep{len12}. 

\begin{figure}
\centering
\subfigure{\includegraphics[width=0.999\textwidth]{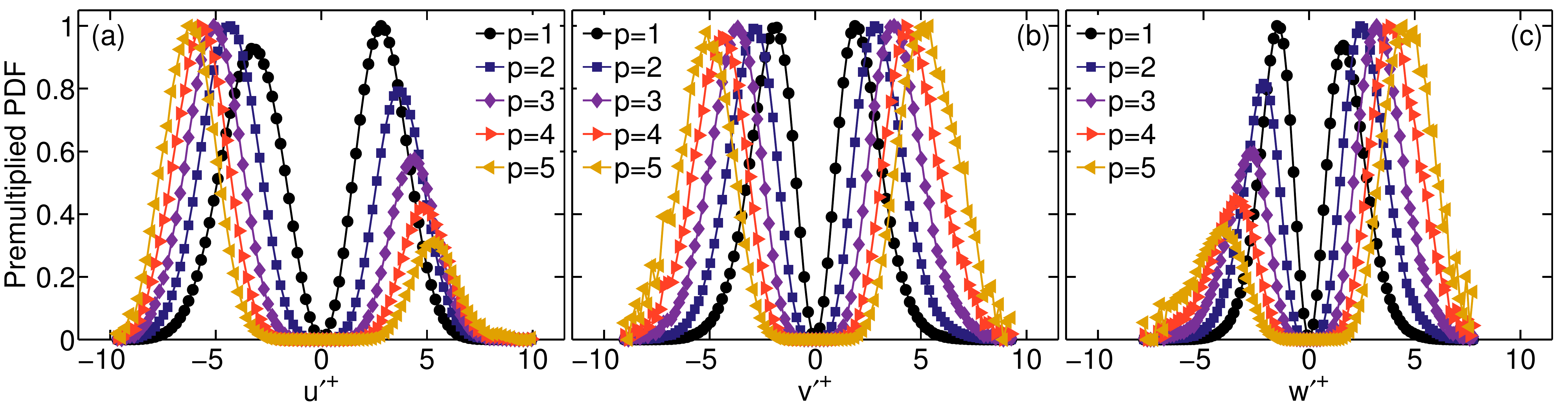}}
\caption{Normalized premultiplied PDFs, i.e.\ ($\mathrm{PDF}(x^{\prime+})(x^{\prime+})^{2p})/(\mathrm{max(PDF}(x^{\prime+})(x^{\prime+})^{2p}))$ where $x$ indicates the (a) streamwise ($u$), (b) spanwise ($v$), and (c) vertical ($w$) velocity component at $z/H=0.1189$ respectively, for $p=1,2,3,4,5$. Each curve has been normalized by its peak for plotting purposes.}
\label{figure5}
\end{figure}

Later on we will  evaluate high-order moments of the fluctuating velocities and thus statistical convergence is an important issue. As done in the analysis of experimental data \citep{men13}, one can test for convergence by examining premultiplied PDFs. Figure \ref{figure5} shows the premultiplied PDFs for the even-order moments up to the $10^\mathrm{th}$-order. The figure shows that the moments, i.e.\ the area under the corresponding curves, can be determined accurately up to the $10^\mathrm{th}$-order moment for the streamwise velocity component. For the spanwise and vertical velocity component we see that the statistics are slightly less converged for the $10^\mathrm{th}$-order moment, but still this convergence can be considered as acceptable.

\section{Results} \label{section3}
In section \ref{section3_1} we compare the streamwise velocity data with experimental data from the Melbourne wind tunnel \citep{hut09,men13}, before discussing the effect of the numerical resolution and near-wall cross-over length scales (from the near wall region to the logarithmic region for the variance) in section \ref{section3_2}. Subsequently we present LES results for the spanwise and vertical velocity components in section \ref{section3_4}.

\subsection{Streamwise velocity component} 
\label{section3_1}

\begin{figure}
\centering
\subfigure[]{\includegraphics[width=0.49\textwidth]{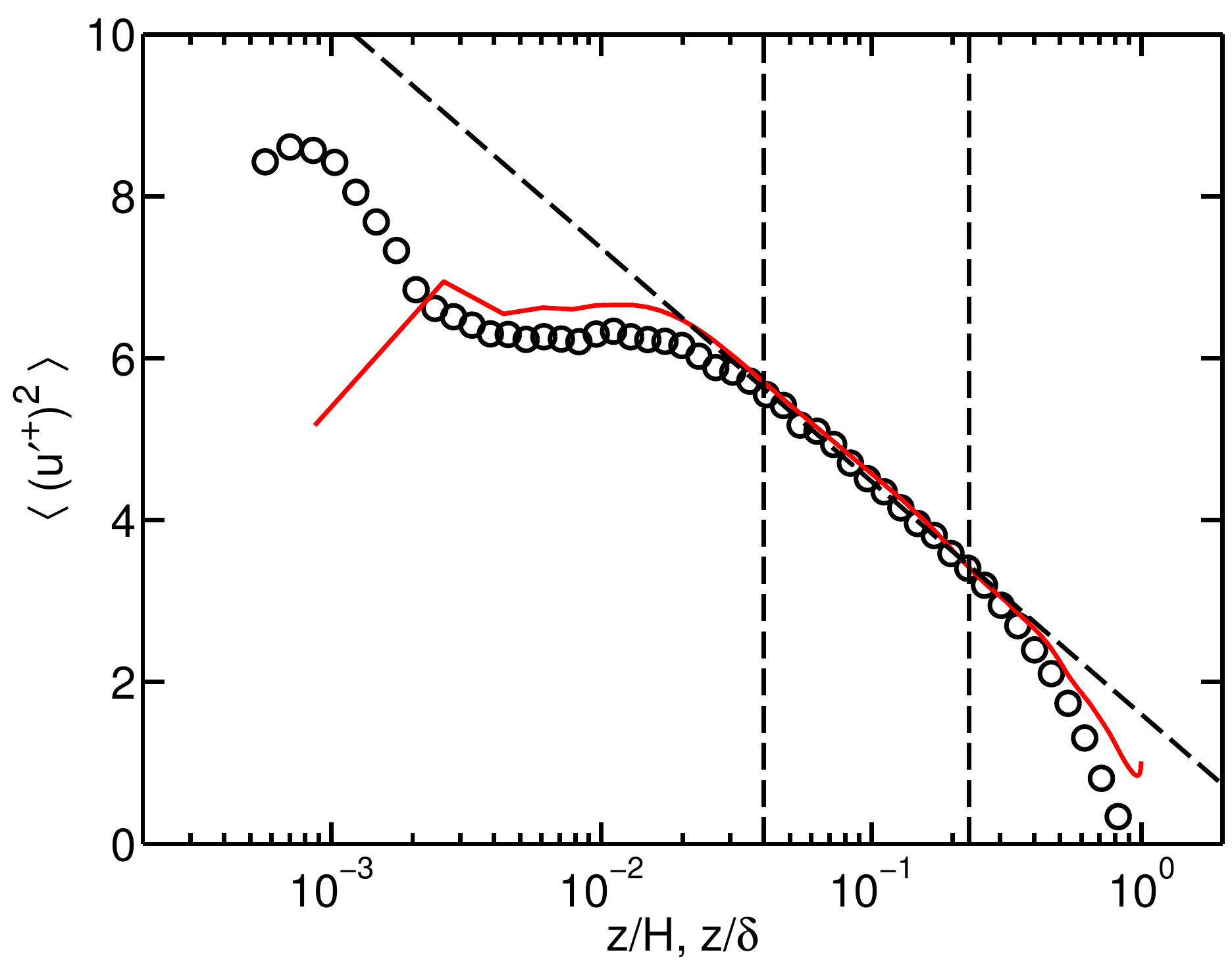}}
\subfigure[]{\includegraphics[width=0.49\textwidth]{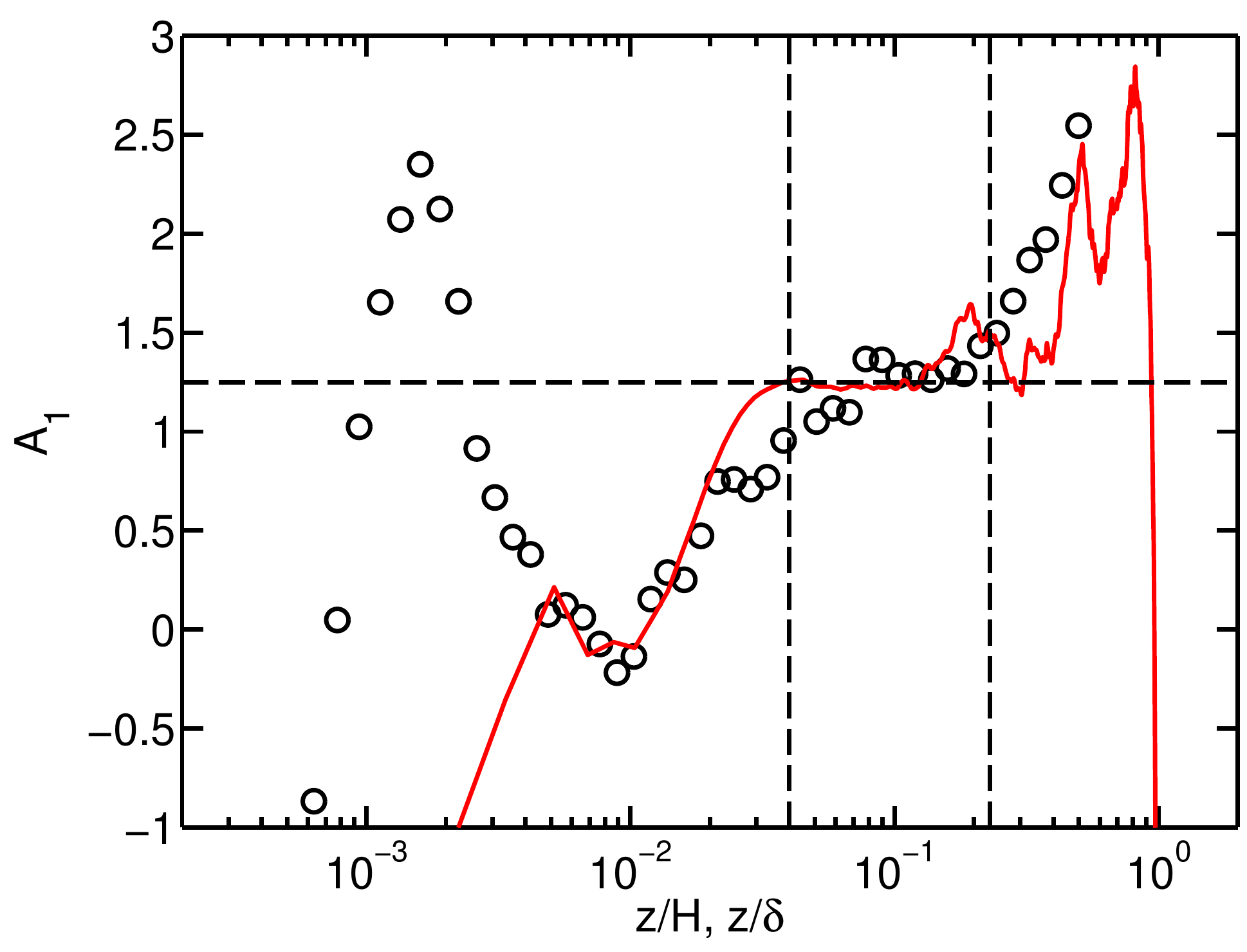}}
\caption{(a) Profile of second-order moment for the streamwise velocity fluctuations obtained from LES (line) compared with the experimental results (circles) \citep{hut09,men13} as function of $z/H$ ($z/\delta$). The dashed line is the fitted logarithmic law for the variance. (b) The local $A_1$ (local slope), see \eqref{equation3}, as function of $z/H$. The dashed vertical lines indicate the region ($0.04 \leq z/H (z/\delta) \leq 0.23$) over which $A_1$ is determined. LES grid locations are  shown in figure \ref{figure2} to \ref{figure4}, figure \ref{figure7} and figure \ref{figure16}.}
\label{figure6}
\end{figure}

Figure \ref{figure6}a shows a comparison of the experimental (empty circles) and LES profiles of variance of streamwise velocity fluctuations. The experiments at $Re_\tau=19030$ from \cite{hut09} are plotted as function of $z/\delta$ where $\delta$ is the boundary layer thickness, while the LES results are plotted as function of $z/H$. For the data, it appears that the equivalence between the two outer scales ($H$ for the LES and $\delta$ for the boundary layer) appears appropriate, since no additional horizontal shifting is seen to be required. The agreement between the LES and the data is quite good for $z/H\gtrsim 0.02$, with a logarithmic layer for the variance visible from about $z/H \approx 0.23$ down to about $z/H \approx 0.02$. This is further confirmed by examining the local slope plot in Fig. \ref{figure6}b, which displays good agreement between LES and experiment down to $z/H \approx 0.02$. The vertical dashed lines indicate the suggested range of the logarithmic region for the variance, and within this region the slope is $A_1 \approx 1.25$ shown as the dashed line in Fig. \ref{figure6}a. 

\begin{figure}
\centering
\subfigure{\includegraphics[width=0.49\textwidth]{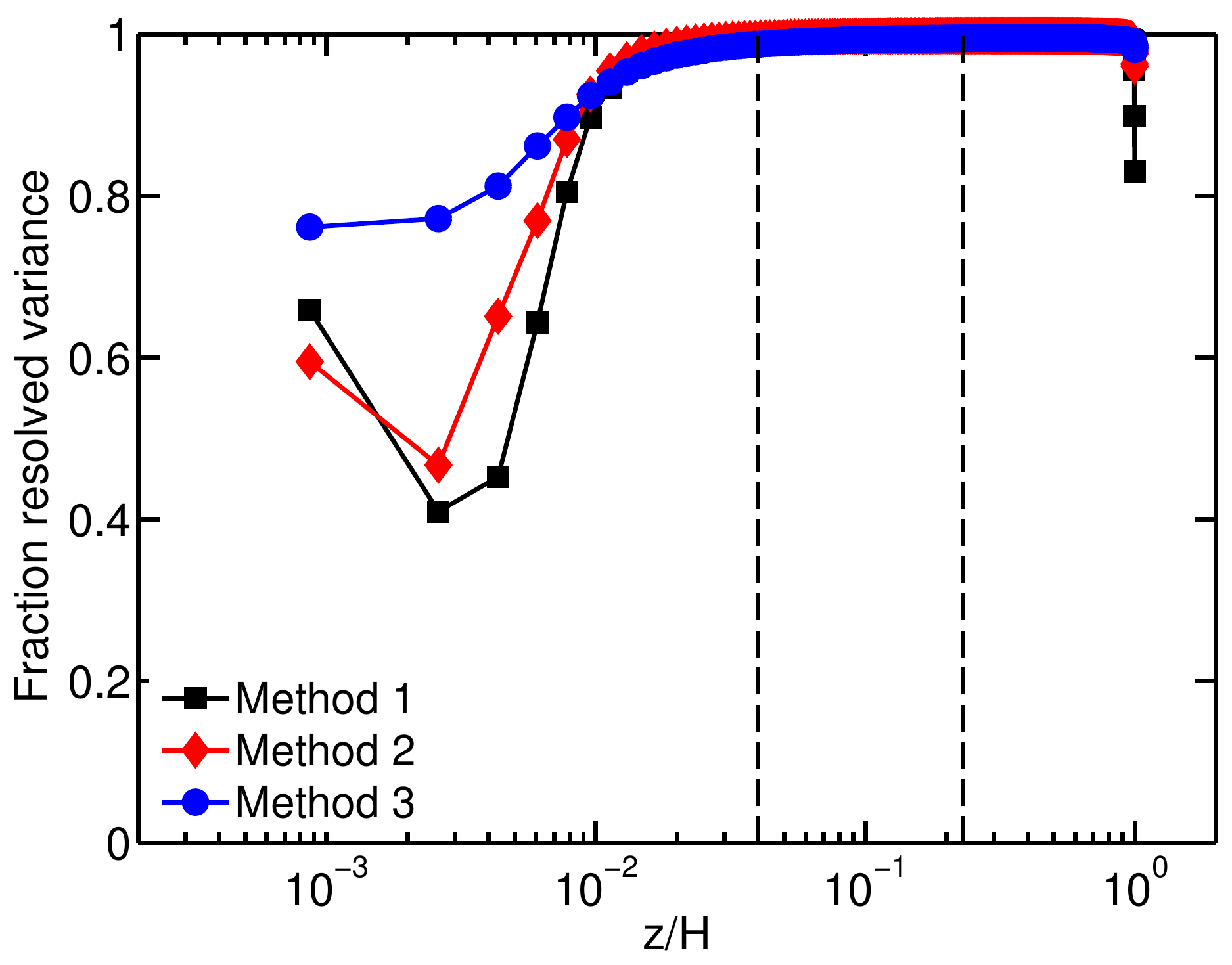}}
\caption{The fraction of the resolved variance of u, $\sigma^2_{\tiny{\mbox{u-LES}}}/\sigma^2_{\tiny{\mbox{u-total}}}$ as function of $z/H$ determined by extrapolation the spectra towards infinity obtained using three different methods, see details in text. The dashed vertical lines indicate the region ($0.04 \leq z/H \leq 0.23$). Closer towards the wall where the fraction of the resolved variance is lower the determination becomes less accurate.}
\label{figure7}
\end{figure}

Note that the reported profiles for the variance from LES corresponds to the `resolved' part of the fluctuations and do not include the sub-grid scale variance. In order to verify that the sub-grid scale variance can be neglected in the logarithmic region for the variance $ 0.04 \leq z/H \leq 0.23$, we examine the variance spectra, see figure \ref{figure3}, in more detail. The second-order moment is related to variance spectra as follows
\begin{equation}
\langle (u'^{+})^2\rangle = \sigma^2_{\tiny{\mbox{u-LES}}} = 2 \int_{0}^{k_{\mathrm{max}}} E_u(k_1;z) u_*^{-2} dk_1,
\end{equation}
where $E_u(k_1;z)$ is the 1-D variance spectrum of the streamwise velocity component in the streamwise wavenumber direction, at height $z$. The integration is over the resolved wavenumber range, from $0$ to $k_{\mathrm{max}}$, where $k_{\mathrm{max}}$ is the largest resolved wavenumber in the LES. By extrapolating the spectra toward infinity the resolved variance in the streamwise velocity component can be compared to its assumed true value. We provide estimates for the total variance by extrapolating the spectrum towards $\infty$ (for practical purposes we approximate infinity with $10000 k_{\mathrm{max}}$) and using 
\begin{equation}
\label{eq_energy1}
\sigma^2_{\tiny{\mbox{u-total}}}  = \sigma^2_{\tiny{\mbox{u-LES}}}  + 2 \int_{k_{\mathrm{max}}}^{\infty} E_{\mbox{\tiny{model}}}(k_1;z) u_*^{-2} dk_1,
\end{equation}
where the second term (the SGS variance, which we will also denote as $\sigma^2_{\tiny{\mbox{u-SGS}}}$) is calculated from an extrapolated spectrum from the LES data. We use three ways to specify $E_{\tiny\mbox{{model}}}$ used in the second term. The first method is a least squares fit to the LES spectrum from $k_{\mathrm{max}}/2$ until $k_{\mathrm{max}}$, the second method a least squares fit to the LES spectrum from $k_{\mathrm{max}}/2$ until $3k_{\mathrm{max}}/4$ and the third method using a $-5/3$ spectrum starting at  $k_{\mathrm{max}}$. The result of this procedure is shown in figure \ref{figure7} and shows that for $ 0.02 \leq z/H \leq 0.95$ more than $98\%$ of the variance in the flow is resolved by our LES, and for most heights it is more than $99\%$. We note that these observations are qualitatively consistent with the observation that the modeled normalized sub-grid scale stresses ($-\langle \tau_{xz}^+ \rangle$) only become larger than $10\%$ of the total stresses ($ -\langle \tau_{xz}^+ \rangle - \langle  u'^{+}w'^{+}  \rangle$) for $z/H \lesssim 0.01$, see figure \ref{figure1}b. Here we emphasize that different methods could be used to estimate the sub-grid scale variance, which could lead to slightly different results. Therefore the mentioned fraction of the resolved variance and the corresponding sub-grid scale variance should be considered as an approximation only. As is shown in figure \ref{figure7} the uncertainty becomes larger closer to the wall due to the difficulties in estimating the sub-grid scale variance when the resolved variance in the flow is only in the order of $50\%$ which happens in the first few grid-points above the wall. Here we also emphasize that the mean and variance profiles obtained from LES are relatively resolution independent in the well resolved region of the flow, while differences are observed closer to the wall where the resolution influences the resolved variance in the flow most.
 
\begin{figure}
\centering
\subfigure[]{\includegraphics[width=0.49\textwidth]{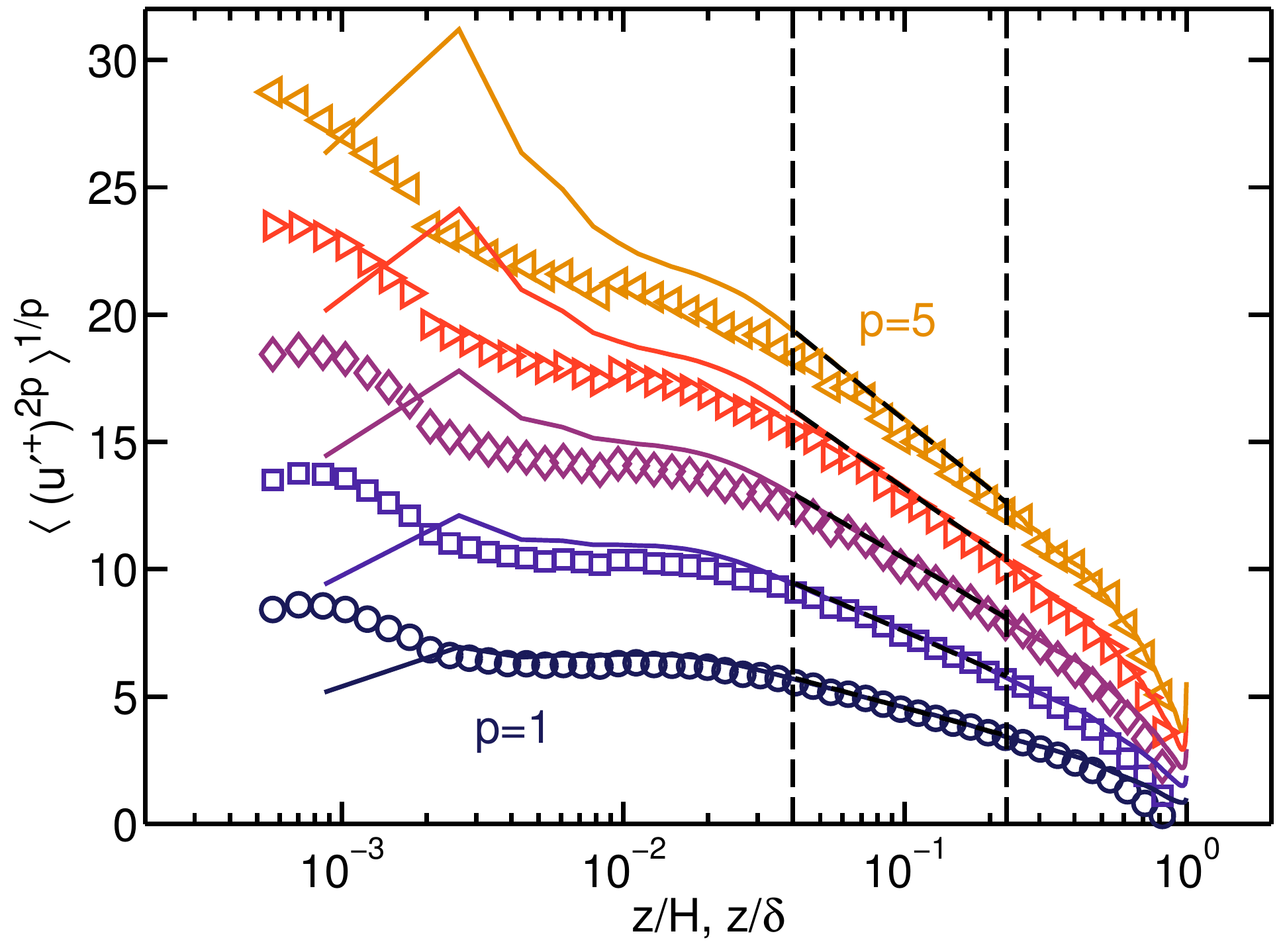}}
\subfigure[]{\includegraphics[width=0.49\textwidth]{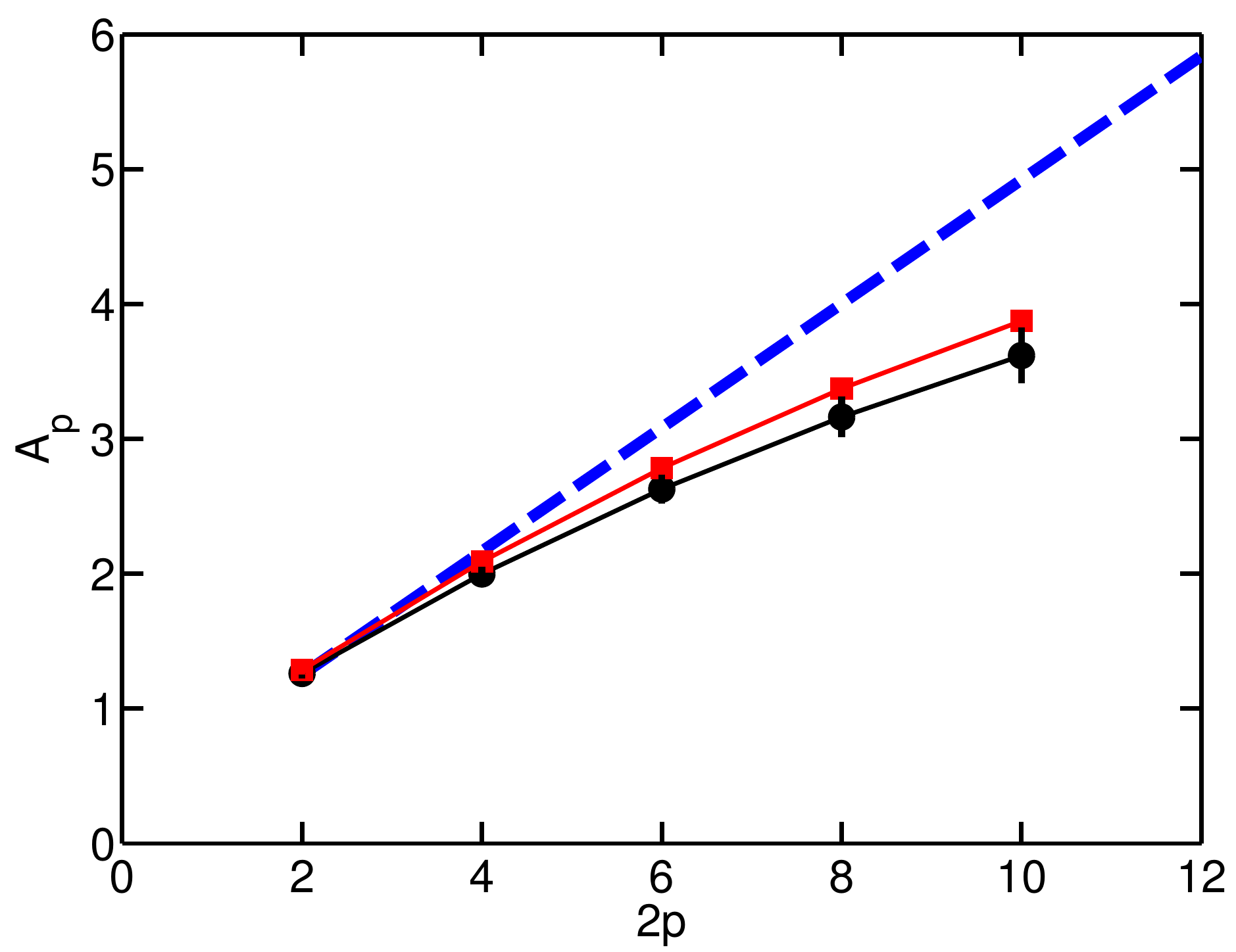}}
\caption{(a) Even-order moments (raised to the $1/p$-th power), of order $2p= 2$ (circles), $4$ (squares), $6$ (diamonds), $8$ (right-pointing triangles), and $10$ (left-pointing triangles) of the streamwise velocity fluctuations as a function of $z/H$ ($z/\delta$). The symbols indicate experimental data \citep{hut09,men13} and the lines LES data. The dashed vertical lines indicate the logarithmic region for the variance ($ 0.04 \leq z/H (z/\delta) \leq 0.23$). (b) $A_p$ as function of $2p$ from the Melbourne experiments at $Re_\tau=19,030$ (circles) and LES (squares). The dashed line indicates the Gaussian values $A_p=A_1 [(2p-1)!! ]^{1/p}$ with $A_1=1.25$. LES grid locations are shown in figure \ref{figure2} to \ref{figure4}, figure \ref{figure7} and figure \ref{figure16}.}
\label{figure8}
\end{figure}

Next, in figure \ref{figure8}a we present profiles of moments of order $2p$, raised to the power $1/p$. It shows that the higher-order, even moments of the streamwise velocity also agree quite well with the experimental data. In agreement with this observation the corresponding $A_p$ values (see \eqref{equation3}) obtained from fitting the data in the interval $ 0.04 \leq z/H \leq 0.23$ also show good agreement, see figure \ref{figure8}b. 

\subsection{Cross-over scale} \label{section3_2}
Estimates for the near-wall start of the logarithmic region for the variance vary significantly. The classical theoretical assumption is that the equilibrium logarithmic layer begins at a fixed value of $z^+$. However, recent experimental evidence \citep{zag98,mar03,hut09,win12,hul12,hut12,mar13} and studies such as those of \cite{wei05}, \cite{eyi08}, and \cite{kle09} suggest that there is a Reynolds-number dependence for the lower limit of this region. Specifically, \cite{kle09}, \cite{alf11} and \cite{mar13} have proposed a $\sim Re_\tau^{1/2}$ dependence for the lower limit of the logarithmic law for the variance. At very high Reynolds numbers, such a trend raises the interesting question about how to represent effects of viscosity in `infinite Reynolds number' LES in which the viscous stress is neglected entirely. Also, the status of such a scaling for rough-wall boundary layers has not been established. Here we examine this issue from the viewpoint of our LES results. In general agreement with observations from experimental data \citep{mar03,hut09,smi11,kul12,win12,hul12,hut12,mar13} we find that the turbulence intensity profiles tend to depart more abruptly from the logarithmic profile for the variance than profiles of the mean velocity when approaching the wall. Figure \ref{figure6} shows the LES data begin to deviate from the logarithmic law for the variance when $z/H \lesssim 0.02$. We point out that also at this height the mean velocity displays a non-negligible ``bump'' as seen in figure \ref{figure2} around the twelfth vertical grid point, i.e.\ $z/H\approx 0.02$. 
 
In the LES, since viscous effects are neglected, there are only a few other characteristic length scales in the near-wall region: the grid resolution $\Delta$ (here we use the simplifying characterization of grid scale as $\Delta = (\Delta x \Delta y \Delta z)^{1/3}$, see \cite{sco93} for a justification), the LES mixing length $C_s \Delta$, where $C_s$ is the dynamically determined Smagorinsky coefficient, and the imposed roughness scale $z_0$. We will denote the height of the break in the variance profiles as $z_b$. One can postulate a simple extension of the two prior models for the lower limit of the logarithmic layer for the variance (fixed $z_{b}^+$, or additional dependence on Reynolds number as $z_b^+ Re_\delta^{1/2}$) to the case of LES in which an `LES inner length-scale' replaces the viscous scale, $\nu/u^*$. Then one is led to four possibilities: that the cross-over scales with grid resolution and then it could occur at either a fixed height $z_b/\Delta$, or at a fixed $\Delta (z_b/\Delta)^{1/2}$ (or with the corresponding mixing length $C_s\Delta$ instead of $\Delta$). Or that the cross-over scales with roughness length $z_0$, again leading to two possibilities: a cross-over at a fixed height $z_b/z_0$ or at a fixed $z_0 (z_b/z_0)^{1/2}$. Naturally, some other powers may be possible, or if $\Delta/H$ is close to $z_0/H$ some intermediate options are possible, including dependencies on both $\Delta/H$ and $z_0/H$.

\begin{figure}
\centering
\subfigure[]{\includegraphics[width=0.49\textwidth]{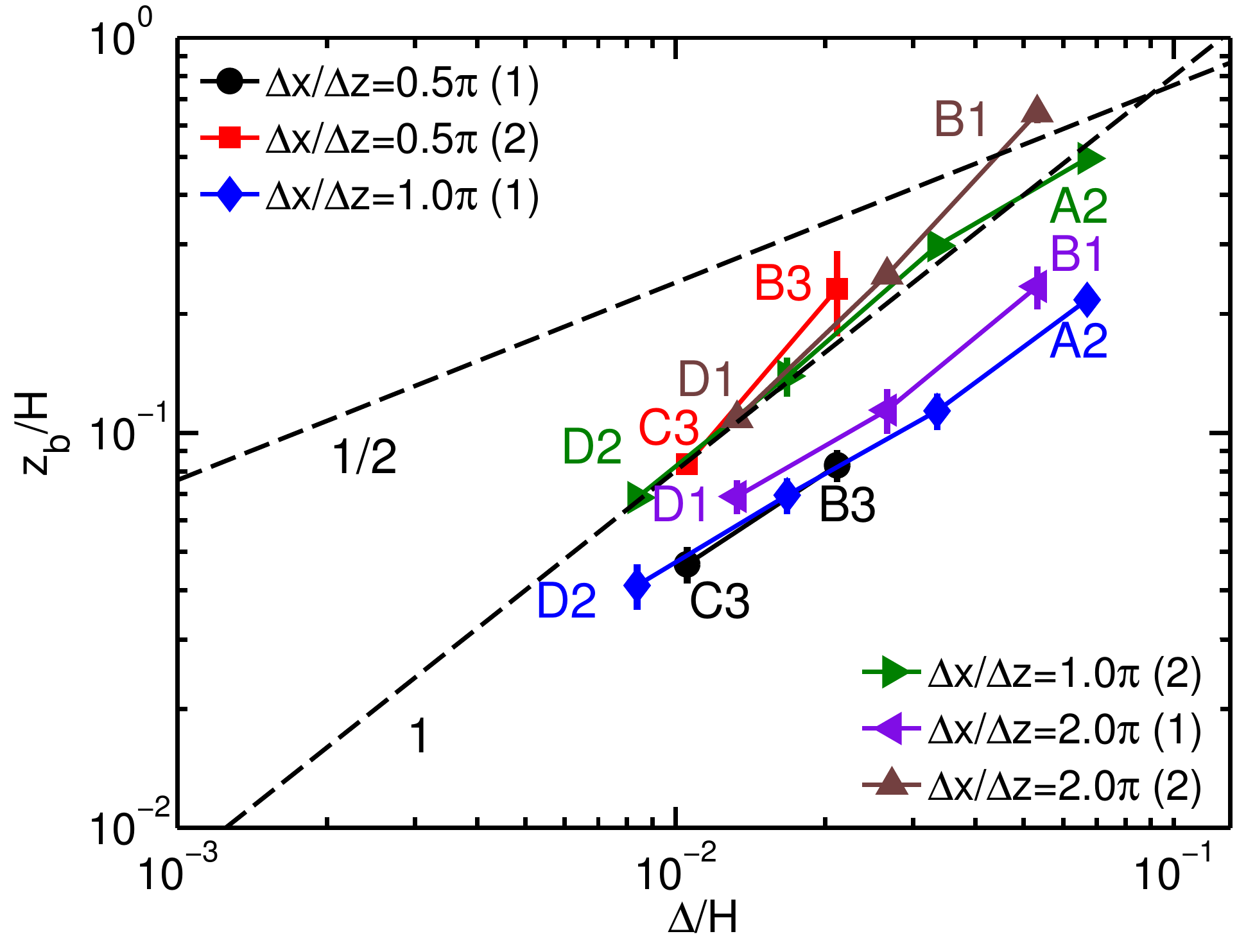}}
\subfigure[]{\includegraphics[width=0.49\textwidth]{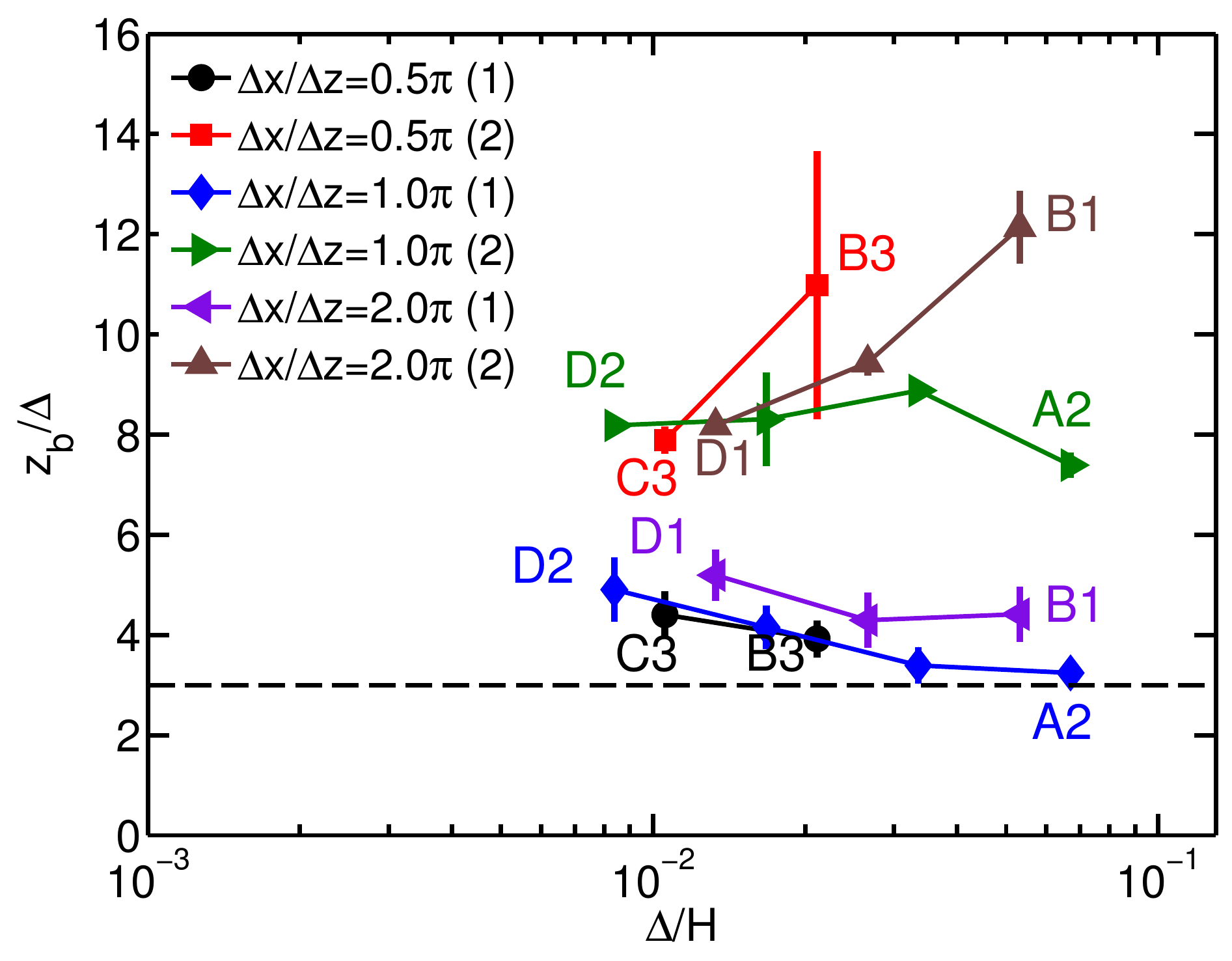}}
\caption{(a) $z_b/H$ as function of $\Delta/H$ compared to $(\Delta/H)^{1}$ and $(\Delta/H)^{1/2}$ behavior (dashed lines). (b) $z_b / \Delta$ as function of $\Delta/H$. The circles, diamonds, and left-pointing triangles indicate $z_b/H$ obtained from $A_1=1$ ((1) in the legend). The squares, right-pointing triangles, and up-pointing triangles indicate $z_b/H$ obtained from $\mathrm{d}A_1 / \mathrm{d}z=0$ ((2) in the legend), see details in the text. The ratio of the horizontal to vertical grid scale is mentioned in the legend, see also table \ref{table1}. }
\label{figure9}
\end{figure}

We first examine the dependence of the cross-over on grid resolution $\Delta$. As discussed before, the spectra (figure \ref{figure3}) and the vertical stress profiles (figure \ref{figure2}b) indicate that a smaller fraction of the variance of the flow is explicitly resolved in this region and the LES modeling therefore becomes more important. The reason for this is that here the horizontal resolution becomes more limiting and at the same time the effect of the modeled wall stresses (see \eqref{equation4}) becomes noticeable in this region. The results indicate that the position at which the LES data for the higher-order moments begin to depart from the logarithmic law depends on the grid resolution. As there is uncertainty in the determination of $z_b$, especially for lower resolution simulations, we find this location using two methods. In the first method we define $z_b/H$ based on the vertical location where $A_1=0.8$, see figure \ref{figure2}b. In the second method $z_b/H$ is based on the position where dA$_1$/dz=0. Figure \ref{figure9} shows $z_b/H$ as function of $\Delta/H$ assuming $z_b>3\Delta$. The $z_b>3\Delta$ criterion is used to prevent that the first local maximum in the $A_1$ as function of $z/H$ profiles, see figure \ref{figure6}b, is identified as the start of the logarithmic region for the variance. For the first method the uncertainty is based on the difference between $z_b$ obtained using $A_1=0.6$ and $A_1=1.0$. For the second method the uncertainty gives the difference in $z_b/H$ obtained using $A_1$ and $A_1$ smoothed over a $3 \Delta z$ interval. The figure shows that the lower boundary at which the logarithmic region for the variance can be observed shifts towards the wall when the grid spacing is decreased. Because the outer boundary of the logarithmic region for the variance occurs at a fixed fraction of the boundary layer thickness (about $0.1 \delta$ to $0.2 \delta$), this means that a sufficient resolution is required to capture the logarithmic region for the variance over a significant interval. For our highest resolution simulation the start of the logarithmic region for the variance is identified to be around $z/H\approx 0.02$ and for some of the lower resolution simulation the logarithmic region for the variance that can be resolved is too small to observe it clearly. Although the data in figure \ref{figure9} show that a small power difference with $(\Delta/H)^1$ dependence cannot be excluded, certainly a scaling with $(\Delta/H)^{1/2}$ does not appear to hold. 

\begin{figure}
\subfigure[]{\includegraphics[width=0.49\textwidth]{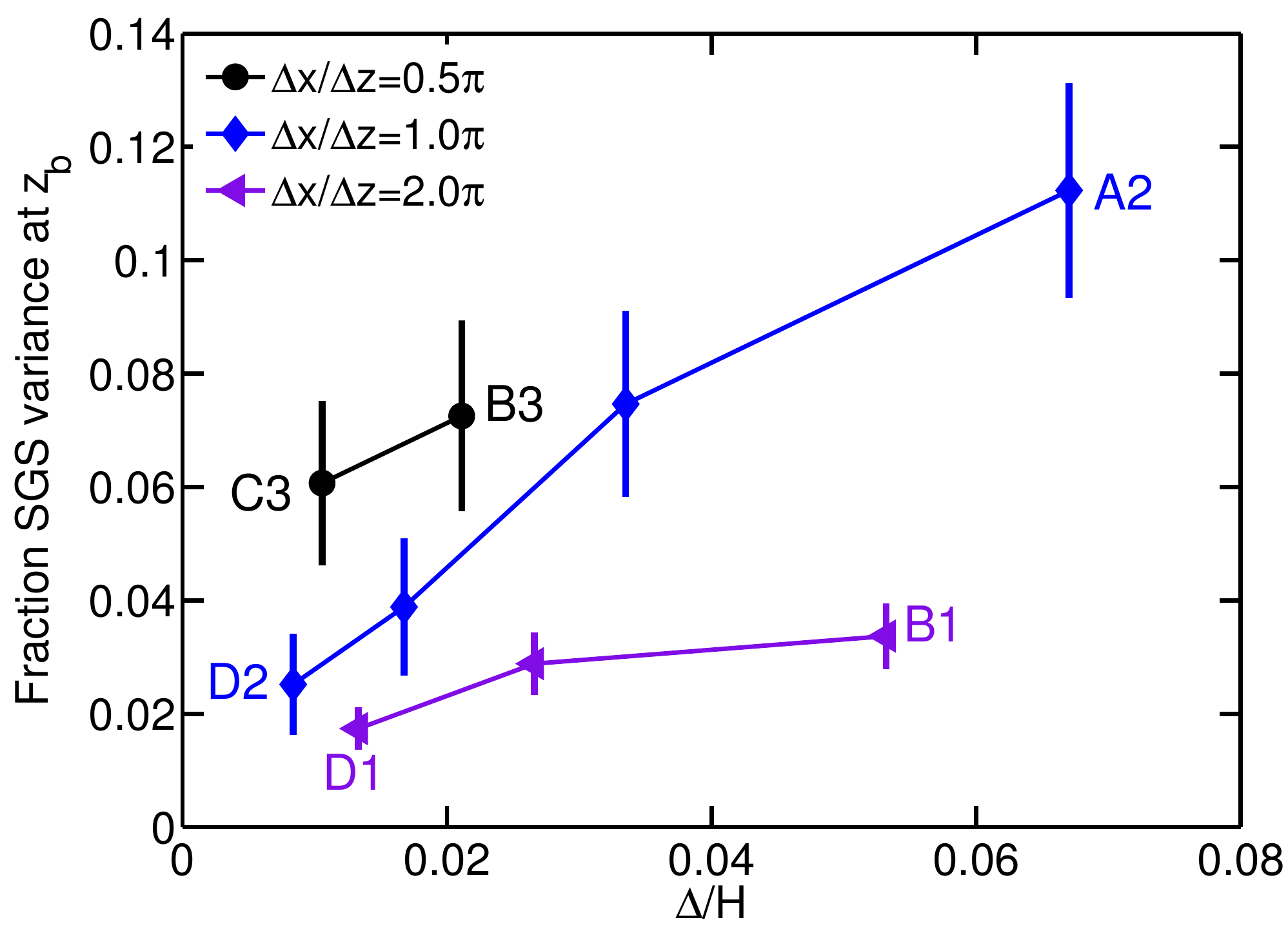}}
\subfigure[]{\includegraphics[width=0.49\textwidth]{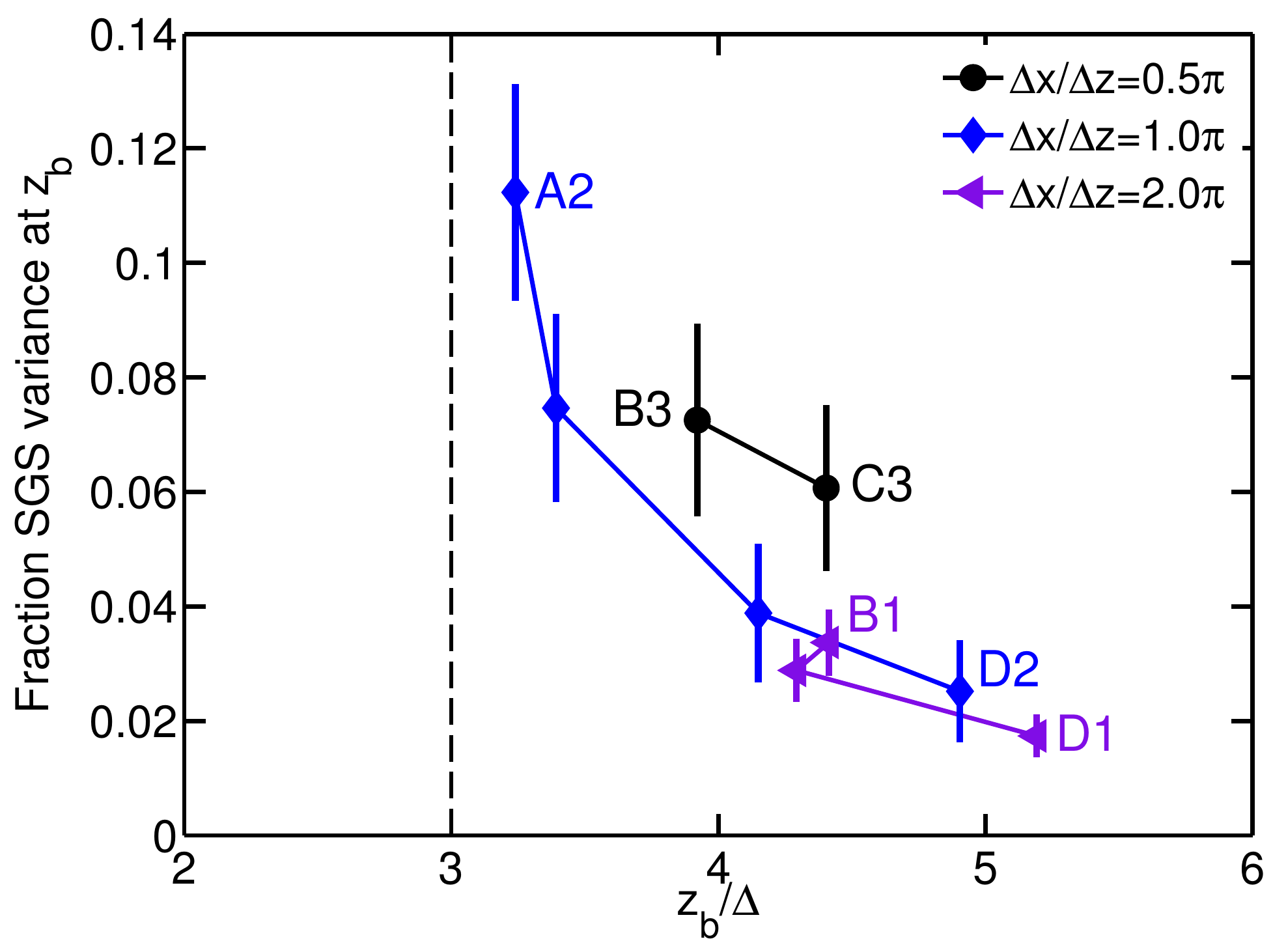}}
\caption{The fraction of SGS variance, $\sigma^2_{\tiny{\mbox{u-LES}}}/\sigma^2_{\tiny{\mbox{u-total}}}$, at the lowest $z_b/H$, see figure \ref{figure9}, as function of (a) $\Delta/H$ and as function of (b) $z_b/\Delta$ using the lower $z_b$ estimates shown in figure \ref{figure9}. Panel a shows that the percentage of the SGS variance decreases with increasing grid resolution (decreasing $\Delta$). Panel b shows that the results nearly collapse with the $z_b/\Delta$ relation found. Note that the $z_b>3\Delta$ criterion is such that the determined $z_b$ value is not directly influenced for the shown results. The vertical bars indicate the variation obtained from the different methods to determine the SGS variance.}
\label{figure10}
\end{figure}

\begin{figure}
\centering
\subfigure[]{\includegraphics[width=0.49\textwidth]{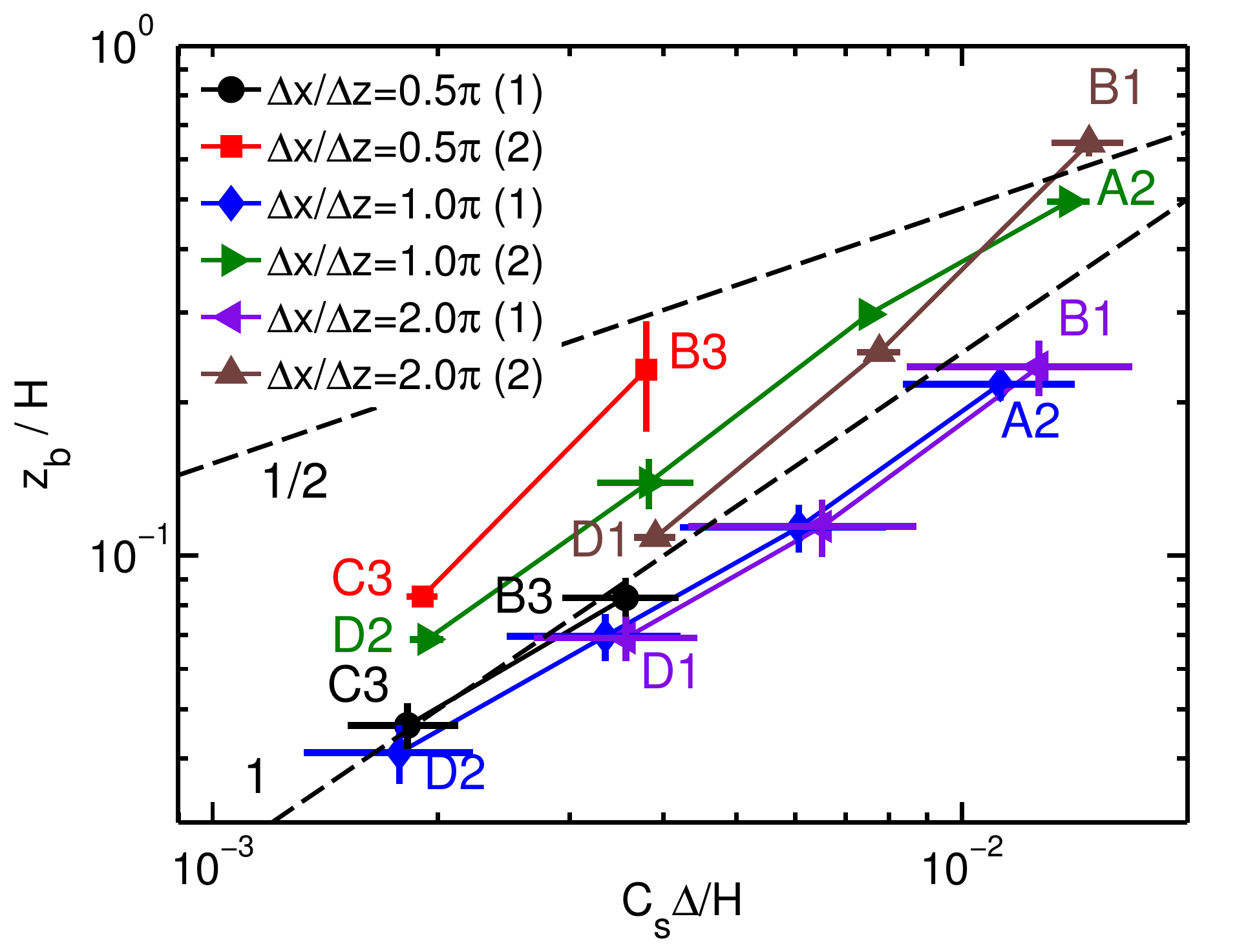}}
\subfigure[]{\includegraphics[width=0.49\textwidth]{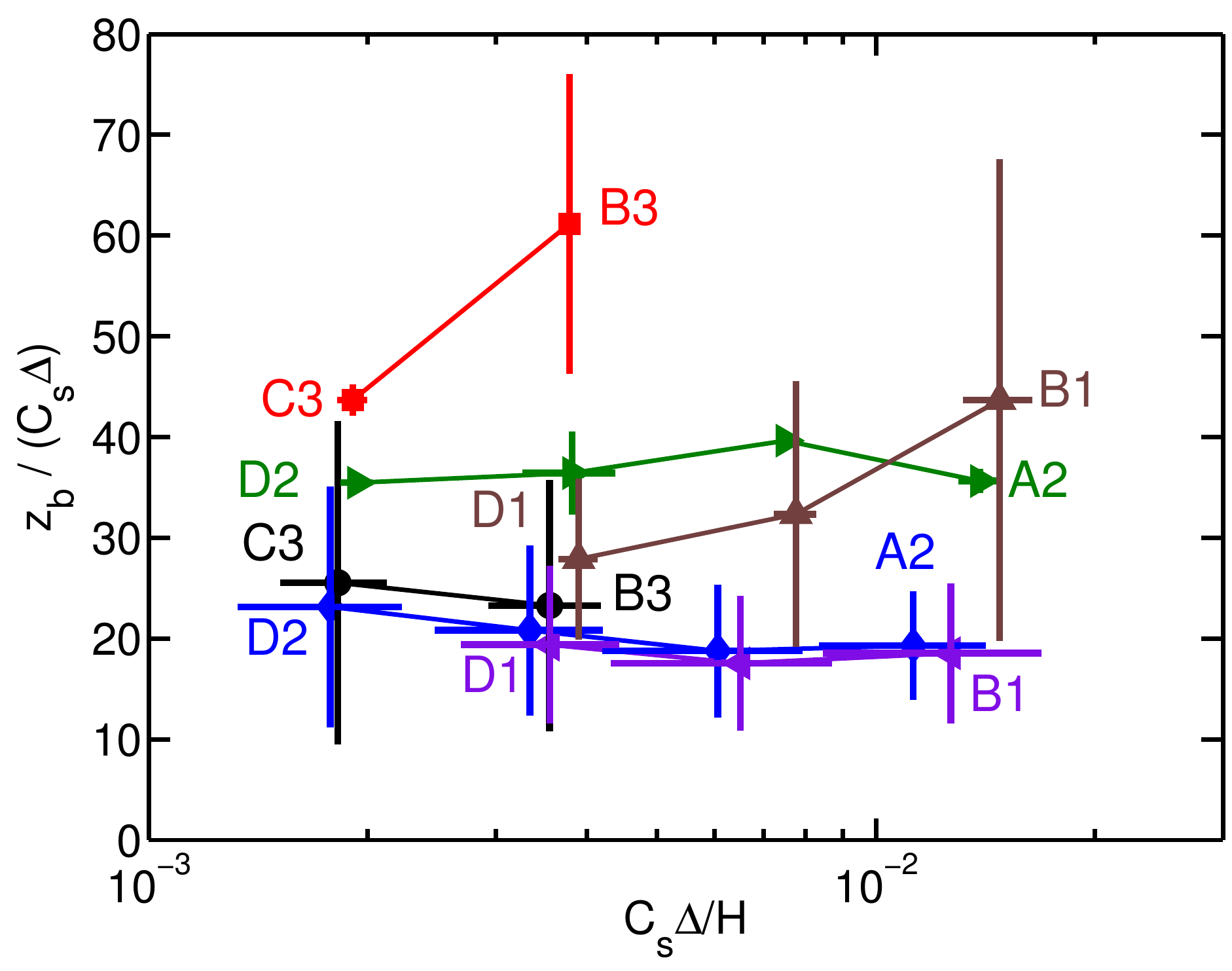}}
\caption{(a) $z_b/H$ as function of $C_s\Delta/H$ compared to $(C_s\Delta/H)^{1}$ and $(C_s\Delta/H)^{1/2}$ behavior (dashed lines). (b) $z_b /(C_s \Delta$) as function of $C_s\Delta/H$. The circles, diamonds, and left-pointing triangles indicate $z_b/H$ obtained from $A_1=1$ ((1) in the legend). The squares, right-pointing triangles, and up-pointing triangles indicate $z_b/H$ obtained from dA$_1$/d$z=0$ ((2) in the legend), see details in the text. The ratio of the horizontal to vertical grid scale, as well as the simulation case numbers as defined in table 1 are indicated. The horizontal and vertical bars indicate the uncertainty determined as indicated in the text.}
\label{figure11}
\end{figure}

Figure \ref{figure10} shows that for all simulations, except for the two lowest resolution simulation ($A1$ and $A2$ (see table \ref{table1}); case $A1$ is omitted from these graphs as $z_b$ cannot be determined well for that simulation) the sub-grid scale variance is less than $10\%$ at $z_b$ and this decreases strongly with increasing resolution. Figure \ref{figure10}b reveals, in agreement with figure \ref{figure9}a, that the results reasonably collapse when represented as function of $z_b/\Delta$. Figure \ref{figure11} shows $z_b/H$ as function of $C_s\Delta/H$ and $z_b /(C_s \Delta$) as function of $C_s\Delta/H$.  As the $C_s$ value at $z_b/H$ is relatively constant, the message of the corresponding figure is similar to the one shown in figure \ref{figure9}, and considering the uncertainty in the data it is hard to say whether $C_s \Delta$ or $\Delta$ length scale is most appropriate.
 
\begin{figure}
\subfigure[]{\includegraphics[width=0.49\textwidth]{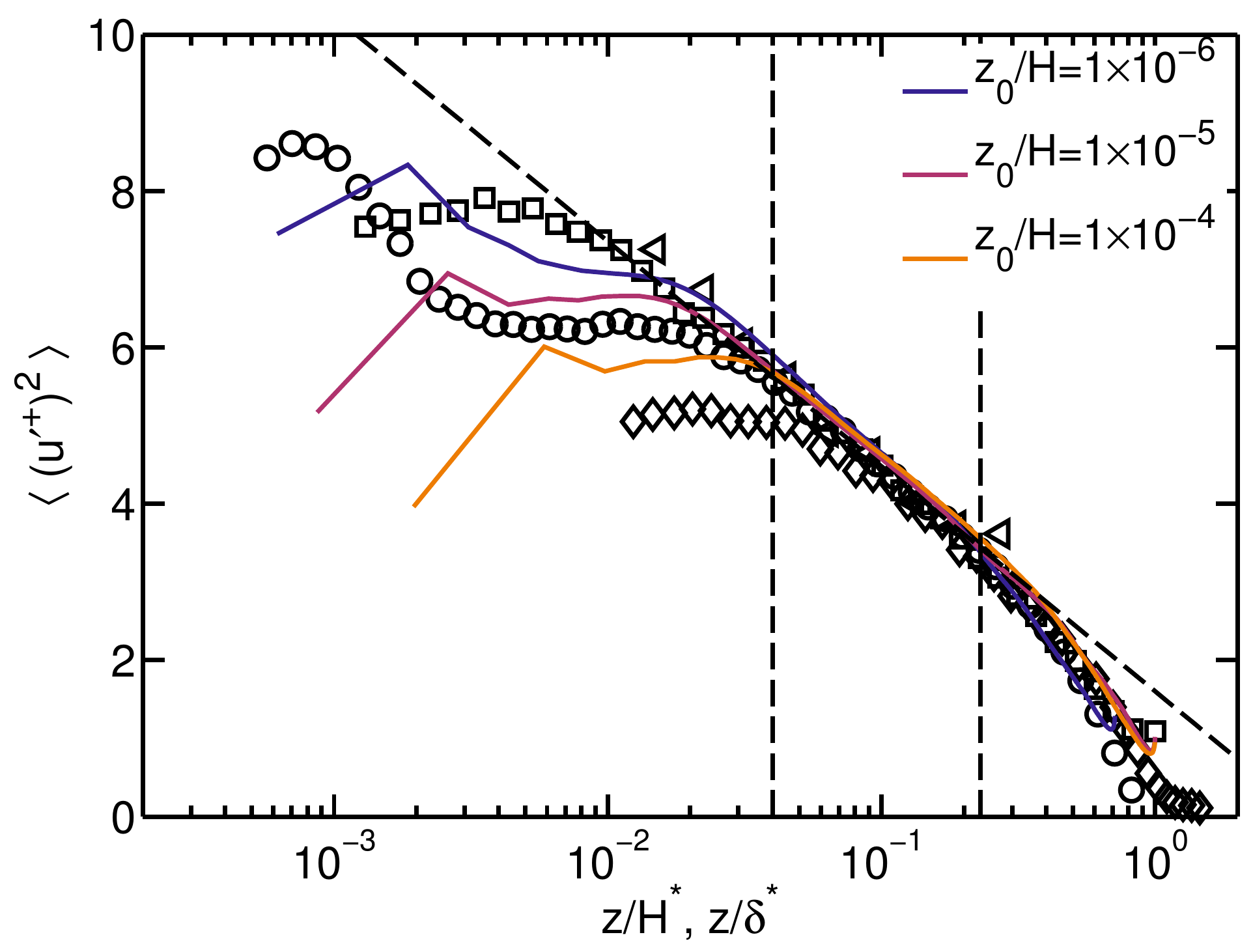}}
\subfigure[]{\includegraphics[width=0.49\textwidth]{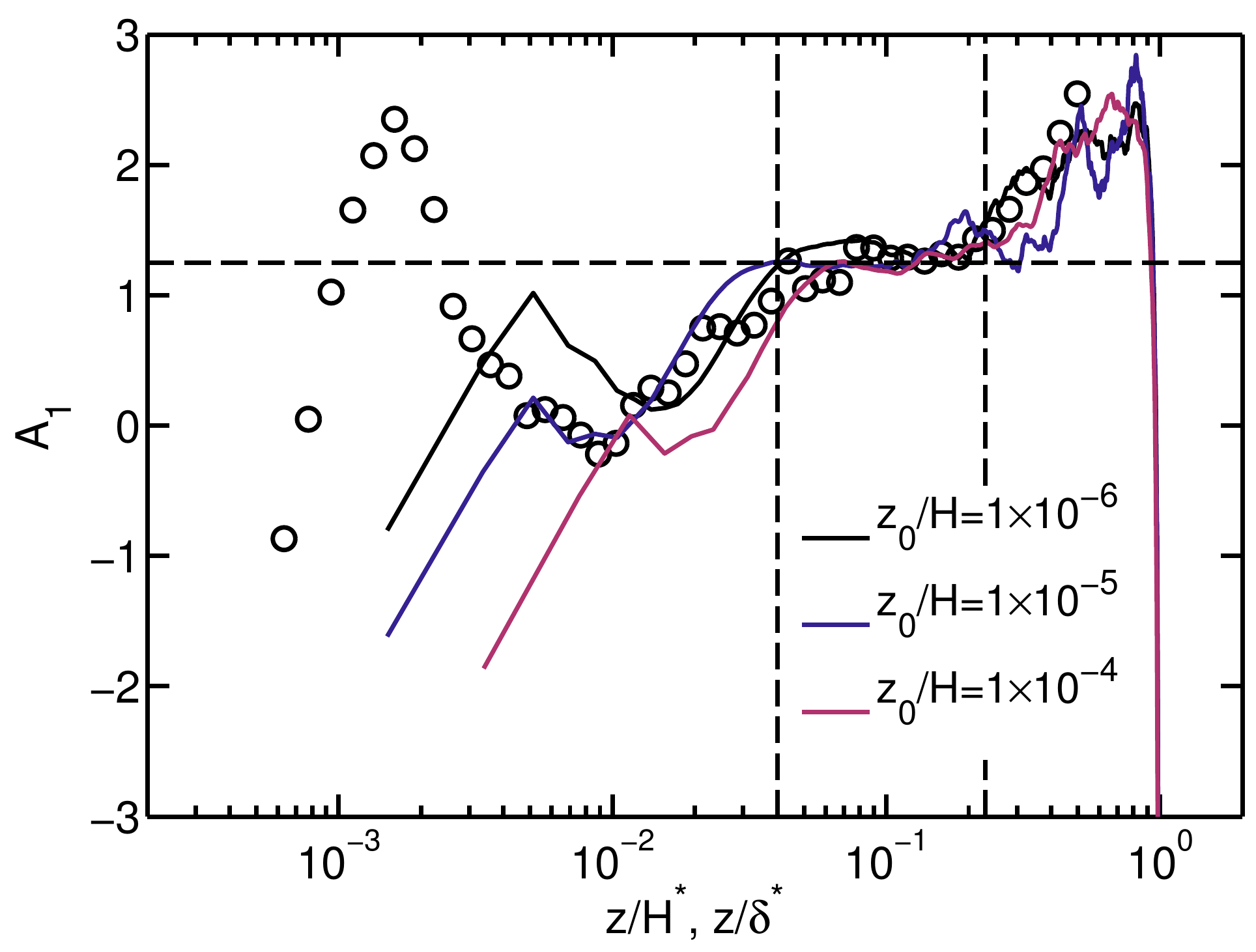}}
\caption{(a) Profile of second-order moment for the streamwise velocity fluctuations obtained from LES compared with the experimental results as function of $z/H^*$ ($z/\delta^*$). (b) The local $A_1$, see \eqref{equation3}. The colors indicate the different $z_0/H^*$ values from the LES. The symbols indicate different experimental data sets: diamonds (roughness, \cite{sch07}), circles (Melbourne, \cite{hut09}), squares (Superpipe, \cite{hul12}), triangles (SLTEST, \cite{hut12}). LES grid locations are shown in figure \ref{figure2} to \ref{figure4}, figure \ref{figure7} and figure \ref{figure16}.} 
\label{figure12}
\end{figure}

\begin{figure}
\subfigure[]{\includegraphics[width=0.49\textwidth]{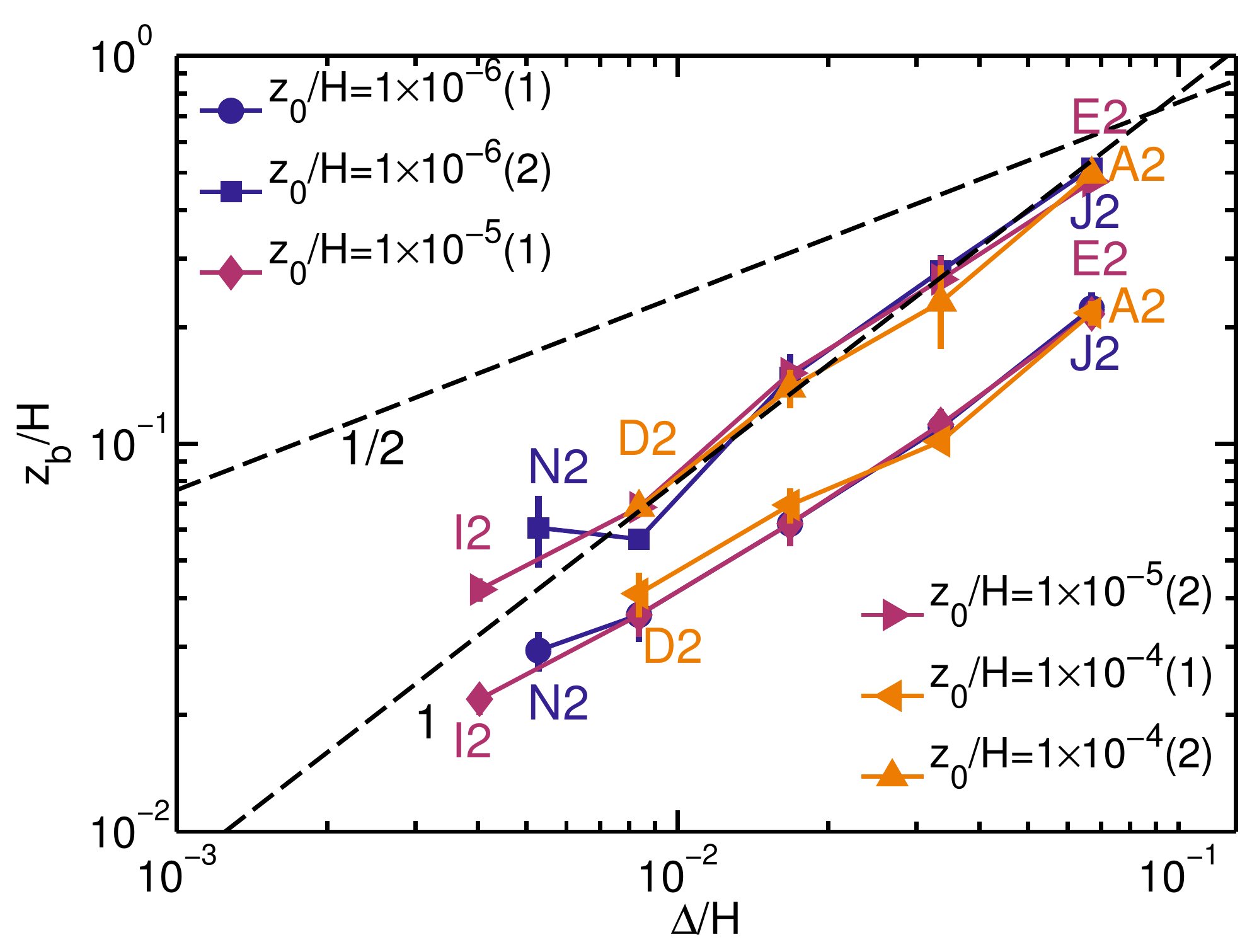}}
\subfigure[]{\includegraphics[width=0.49\textwidth]{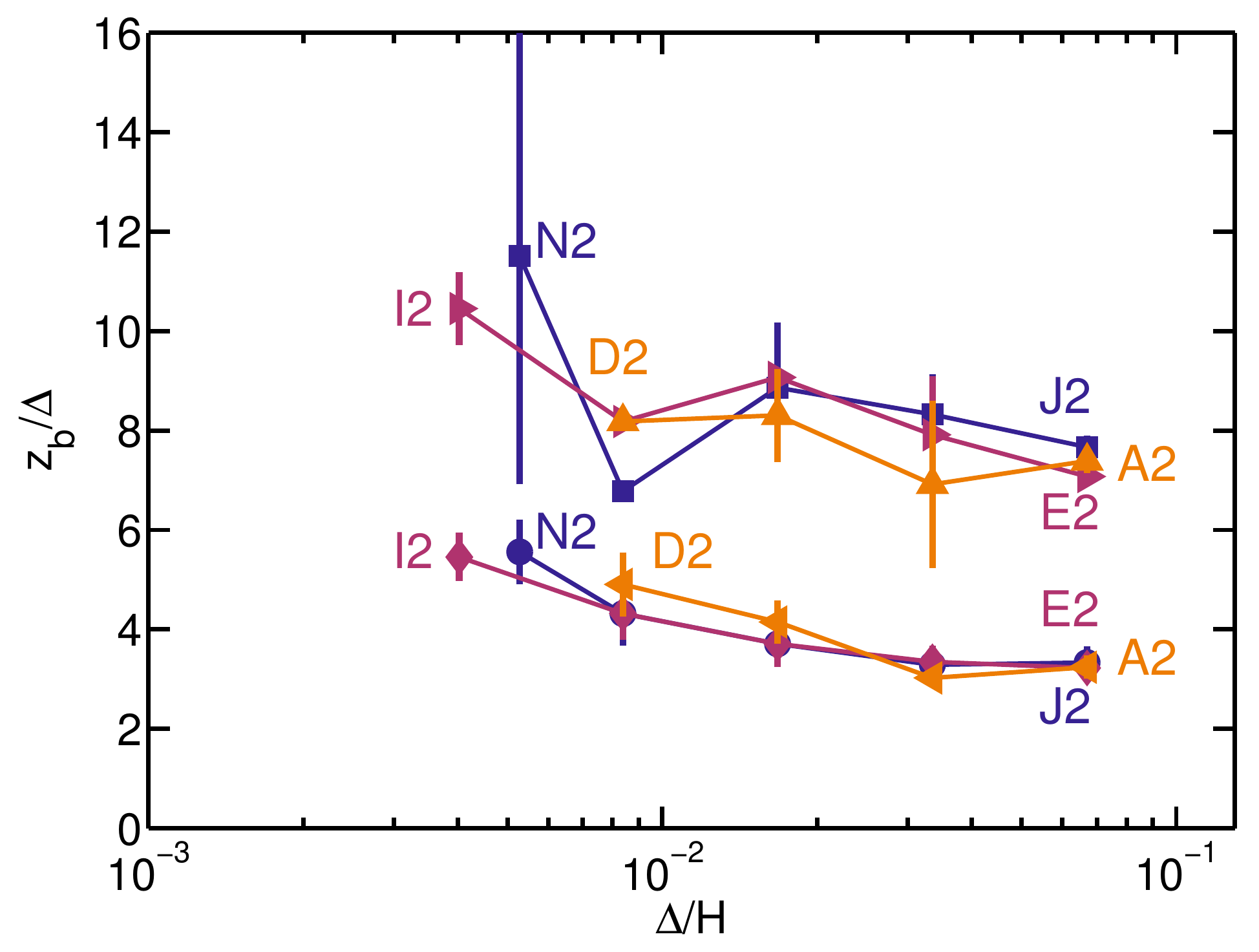}}
\caption{(a) $z_b/H$ as function of $\Delta/H$ compared to $(\Delta/H)^{1}$ and $(\Delta/H)^{1/2}$ behavior (dashed lines) plotted for various $z_0$ to quantify dependence on $z_0$. (b) $z_b / \Delta$ as function of $\Delta/H$. The circles, diamonds, and left-pointing triangles indicate $z_b/H$ obtained from $A_1=0.9$ ((1) in the legend). The squares, right-pointing triangles, and up-pointing triangles indicate $z_b/H$ obtained from $\mathrm{d}A_1 / \mathrm{d}z=0$ ((2) in the legend), see details in the text. The $z_0/H$ value is indicated in the legend, see also table \ref{table1}. }
\label{figure13}
\end{figure}

\begin{figure}
\subfigure[]{\includegraphics[width=0.49\textwidth]{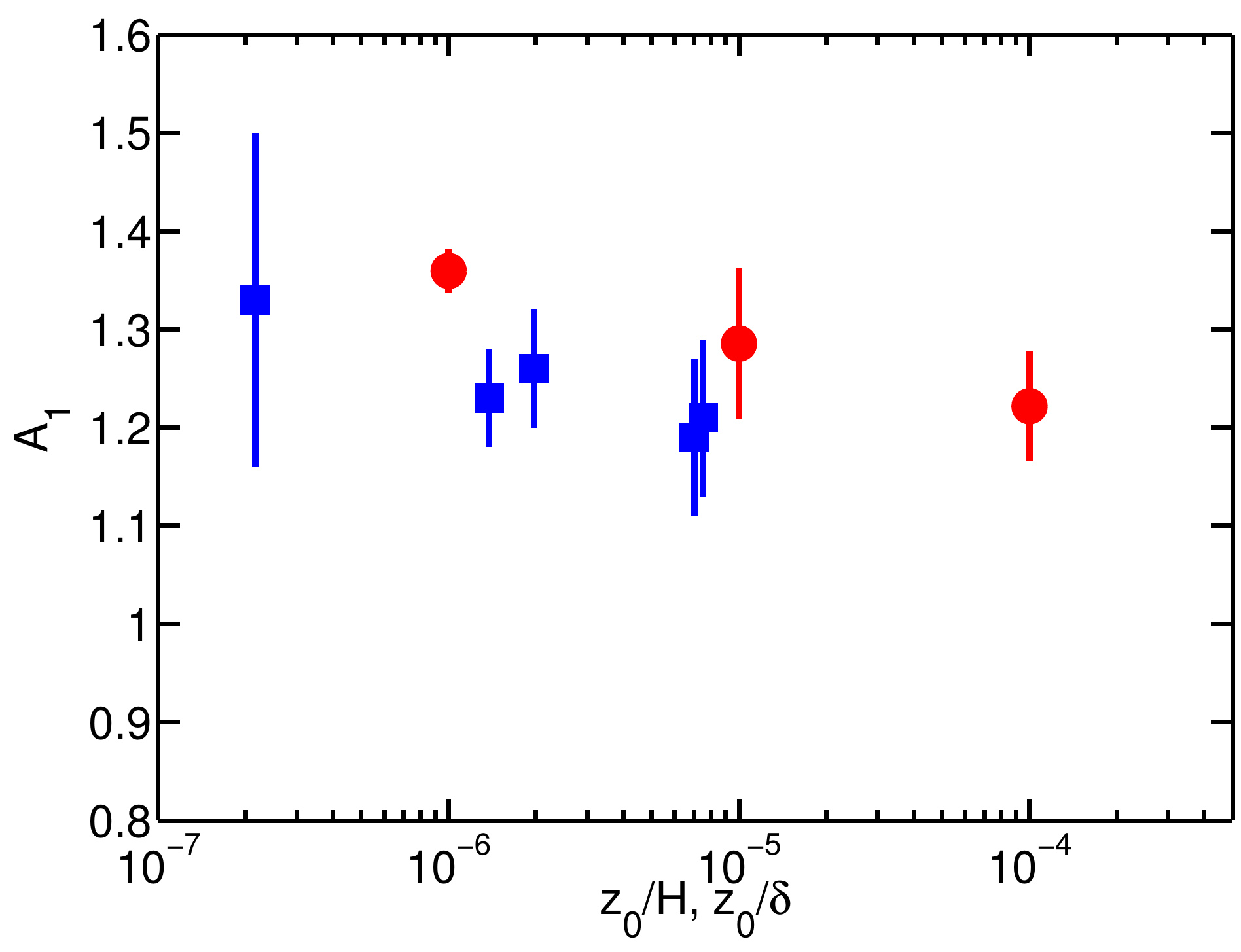}}
\subfigure[]{\includegraphics[width=0.49\textwidth]{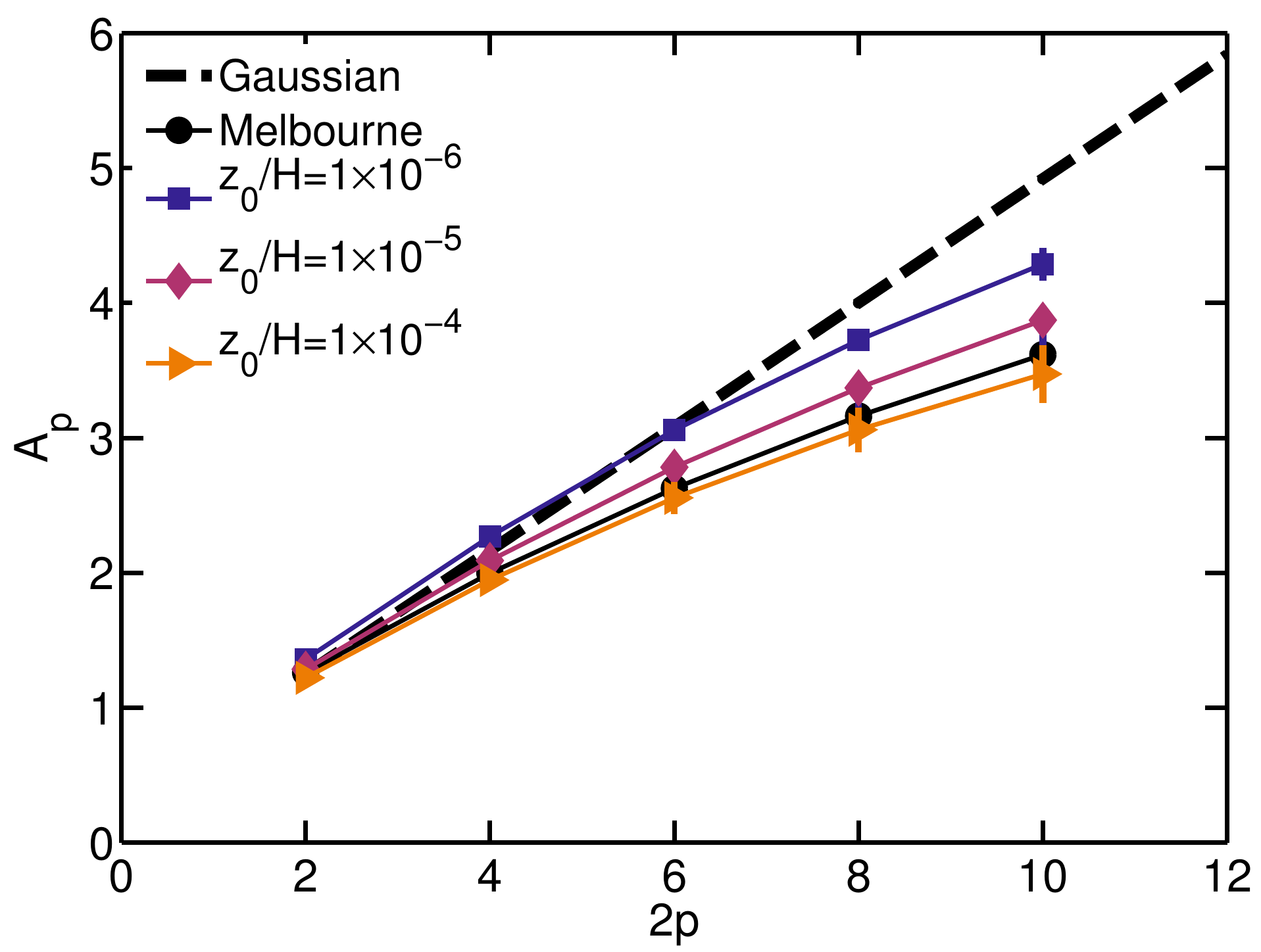}}
\caption{(a) $A_1$ as function of $z_0/H$ ($z_0/\delta$) for different experiments (squares, \cite{mar13}) and the LES (circles). Note that for the experimental data $z_0/\delta$ is obtained from (\ref{eq:z0nu}) as described in the text. (b) $A_p$ as function of $2p$ from experiments (circles) and LES with the different $z_0/H$. The dashed line indicates the Gaussian comparison $A_p=A_1 [(2p-1)!! ]^{1/p}$ with $A_1=1.25$.}
\label{figure14}
\end{figure}

Next, we examine the possible dependence of the cross-over height $z_b/H$ on the imposed roughness scale $z_0/H$. Figure \ref{figure12} shows the second-order moments of fluctuating streamwise velocity for the simulations and several experiments with the different $z_0/H^*$ and $z_0/\delta^*$, and the corresponding local $A_1$. Here $H^*$ and $\delta^*$ are chosen such that all the shown data sets overlap in order to focus on $A_1$ and not on $B_1$, which is discussed below. From figure \ref{figure13} we conclude that $z_b/H$ is roughly independent of $z_0/H$. Simulations using larger roughness lengths (i.e. $z_0/H=2\times10^{-4}$, not shown) and at high resolution suggest that when $z_0/H$ is no longer much smaller than $\Delta/H$, the assumptions on which the equilibrium wall boundary condition is based, begin to lose validity and results (not shown) are degraded. In figure \ref{figure12}a one can also notice that the fluctuations close to the wall decrease with increasing roughness. The reason is that the rougher surface will result in a larger damping of the streamwise velocity fluctuations close to the wall. In figure \ref{figure12}b we see that this results in a slight decrease of $A_1$ as function of the roughness length. We note that the differences in the local $A_1$ obtained for the different roughness lengths in the LES are mainly due to the difference in the resolutions used for these cases. 

Figure \ref{figure14}a shows $A_1$ obtained from the region $ 0.04 \leq z/H \leq 0.23$ compared to the experimental data from several high $Re$ number experiments as summarized in table 1 of \cite{mar13}. In order to relate the inner scale between our LES and the experimental data on smooth wall boundary layers we use the approximate relationship 
\begin{equation}\label{eq:z0nu}
\frac{z_0}{\delta}= \frac{\nu}{\delta u_*} \exp(-\kappa B),
\end{equation}
where $\nu$, $u_*$ and $\delta$ are the air viscosity, friction velocity and boundary layer height in the experiment. The empirical values $\kappa=0.4$ and $B=5$ and the $\nu$, $u_*$ and $\delta$ values as documented in table 1 of \cite{mar13} are used.

As the $B_1$ value in the logarithmic law for the variance depends on the large-scale flow geometry (e.g. it is expected to differ for channels and boundary layers), and because we are mainly interested in capturing the ``universal" slope $A_1=1.25$ behavior, we show the data in figure \ref{figure12}a as function of $z/H^*$. For the experiments the uncertainties shown as error bars are the ones given in table $1$ and $2$ of \cite{mar13}. For the LES we determine the uncertainty in the same way as done for the experimental data (\cite{mar13}) by determining the $95\%$ confidence bounds from the curve-fitting procedure. In order to obtain values consistent with the experimental ones we interpolate the LES data to the measurements heights used in the experiments. The figure suggests that $A_1$ slowly decreases with increasing roughness although the trend is weak compared to the uncertainties. We also recall the observation of \cite{men13} that $A_p$ for the higher-order moments becomes less sensitive to $Re_\tau$ for increasing Reynolds number. Additional experimental and simulation results are needed to verify whether an actual $z_0$ dependence exists. 

\subsection{The role of $B_1$}
In contrast to the fairly constant $A_1$ value it has been shown by \cite{mar13} that $B_1$ can vary significantly among different flows, indicative of dependencies on non-universal large-scale structures in turbulent wall-bounded flows. The data in figure \ref{figure14} show the observed variation of $B_1$ as function of $z_0/H$. The figure shows that $B_1$ obtained from LES is within the scatter obtained from the experiments. Since the half-channel flow geometry in our LES differs from the developing boundary layer experiments, to the degree that 
there are differences, these are to be expected. One can notice that the the $B_1$ value obtained from LES is higher for $z_0/H=10^{-6}$ than for the other two LES cases. We are not sure what the reason is for this difference. The $z_0/H=10^{-6}$ case is the more challenging case since its lower roughness leads to higher mean velocities (compared to the friction velocity). As the increase in the velocity fluctuations is small compared to the increase in the mean velocity, the turbulence intensity is significantly lower for this case than for the other cases. Therefore a larger computational domain to prevent unphysical streamwise and spanwise correlations associated to the use of periodic boundary conditions is necessary. A too small domain leads to higher fluctuations. In addition, the $\Delta/z_0$ and $(\Delta_x=\Delta_y)/\Delta_z$ are largest for $z_0/H=10^{-6}$, which could influence the results. However, unfortunately it is at the moment not possible to perform the $z_0/H=10^{-6}$ with the same $\Delta/z_0$ and $(\Delta_x=\Delta_y)/\Delta_z$ as the other cases since this would require grids with more than an order of magnitude more grid points. 

\begin{figure}
\centering
\subfigure[]{\includegraphics[width=0.49\textwidth]{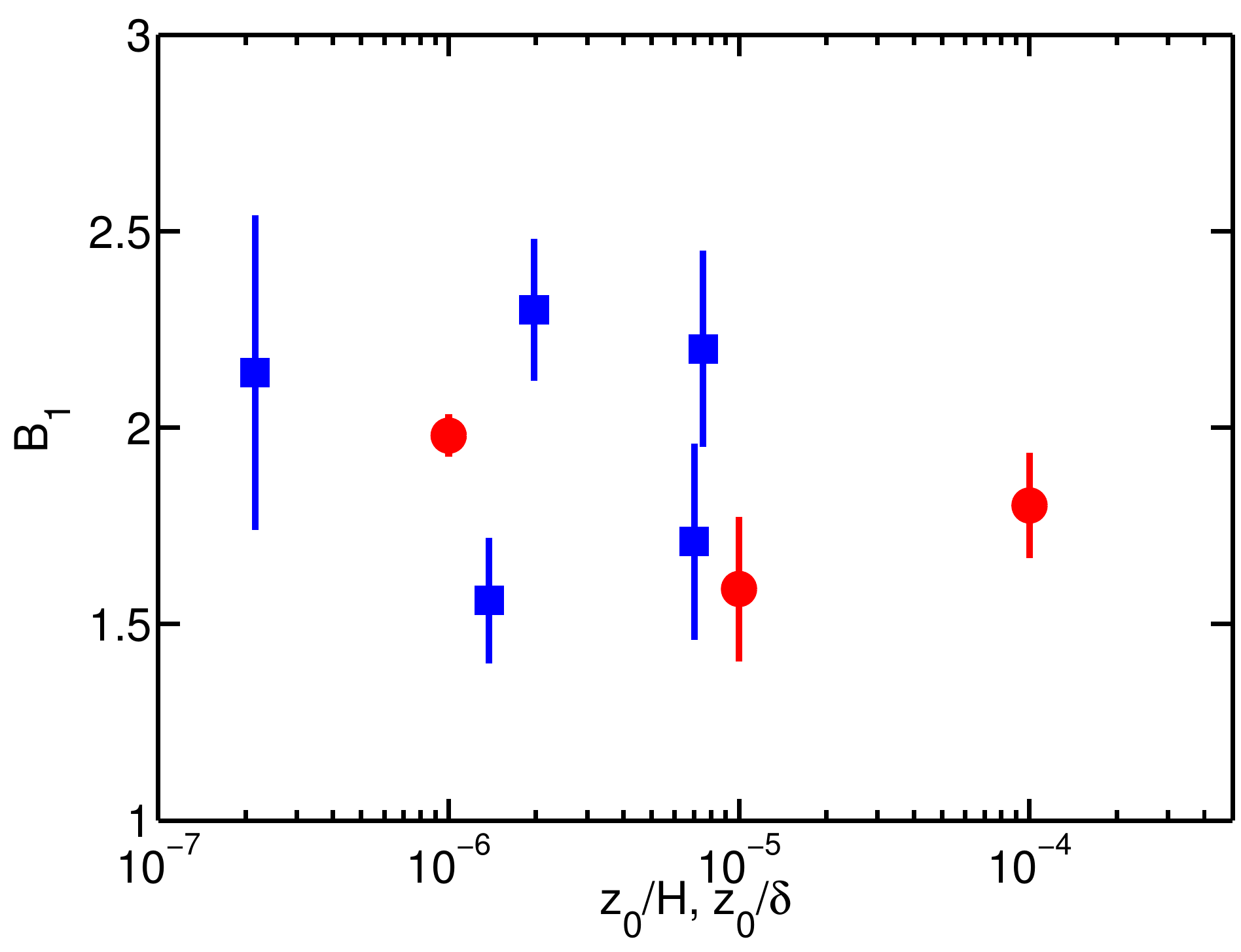}}
\subfigure[]{\includegraphics[width=0.49\textwidth]{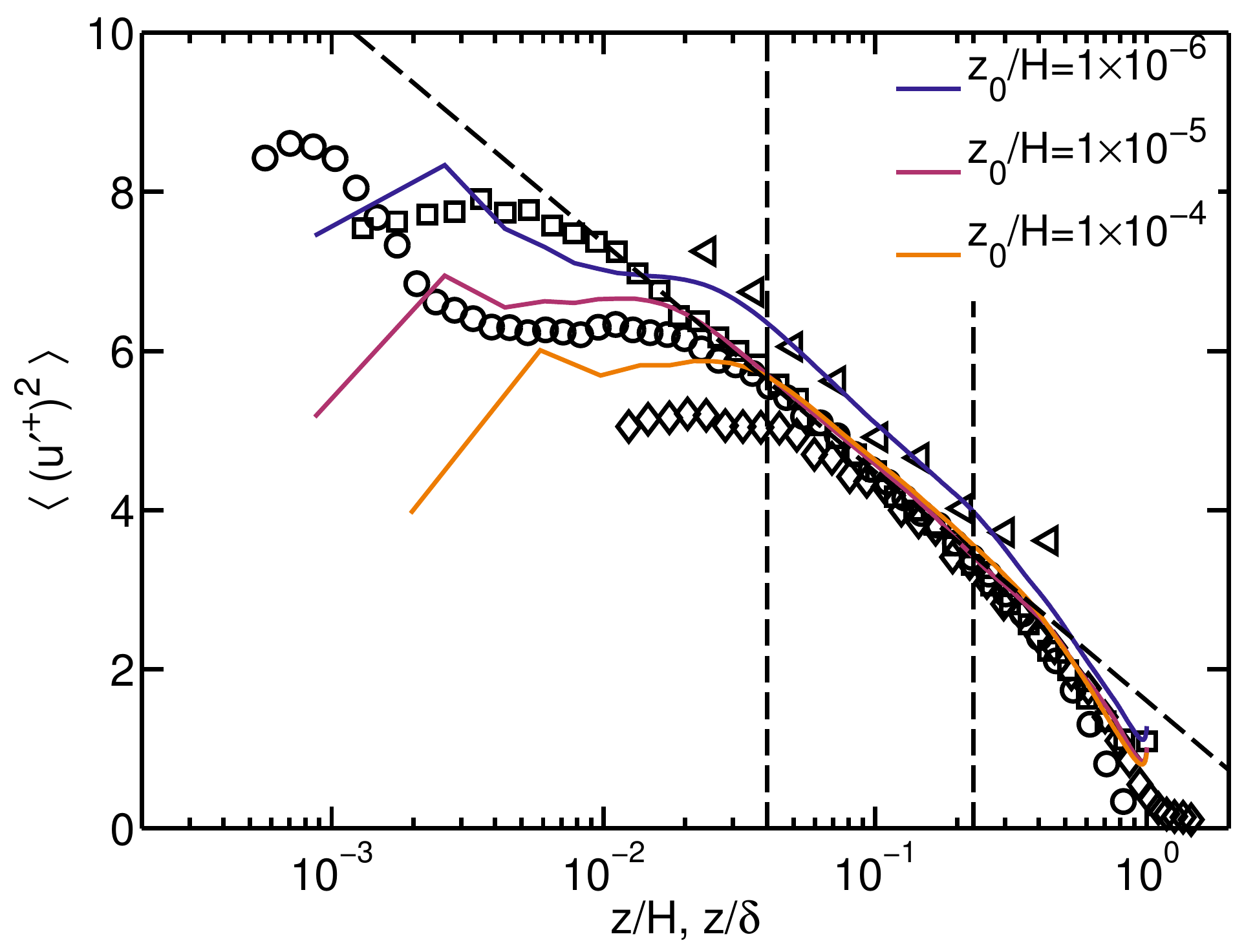}}
\caption{(a) $B_1$ as function of $z_0/H$ ($z_0/\delta$) for different experiments (squares, \cite{mar13}) and the LES (circles). Note that for the experimental data, $z_0/\delta$ is obtained from (\ref{eq:z0nu}) as described in the text. (b) Symbols indicate different experimental data sets: diamonds (rough wall boundary layer, \cite{sch07}), circles (Melbourne, \cite{hut09}), squares (Superpipe, \cite{hul12}), triangles (SLTEST, \cite{hut12}). LES grid locations are shown in figure \ref{figure2} to \ref{figure4}, figure \ref{figure7} and figure \ref{figure16}.}
\label{figure15}
\end{figure}

\subsection{Spanwise and normal velocity components} \label{section3_4}
We now turn to the fluctuations of the spanwise velocity component. Figure \ref{figure15}a shows the higher-order moment data obtained from LES for the spanwise fluctuations. The data in this figure do not reveal as clear a logarithmic region for the variance as the streamwise component results. As indicated before, the spanwise velocity fluctuations are more difficult to resolve than the streamwise velocity fluctuations since their characteristic length scales tend to be smaller than the elongated ones in the streamwise direction. This probably means that the results for the spanwise velocity component are more sensitive to the numerical resolution than for the streamwise velocity component. As we have seen in the previous section that the logarithmic region for the variance of the streamwise velocity can only be captured clearly when the grid is sufficiently fine, we cannot exclude that LES at significantly higher spanwise resolution could reveal a logarithmic region for the spanwise velocity fluctuations as well, but we believe this observation cannot be made based on the current dataset.

\begin{figure}
\centering
\subfigure[]{\includegraphics[width=0.49\textwidth]{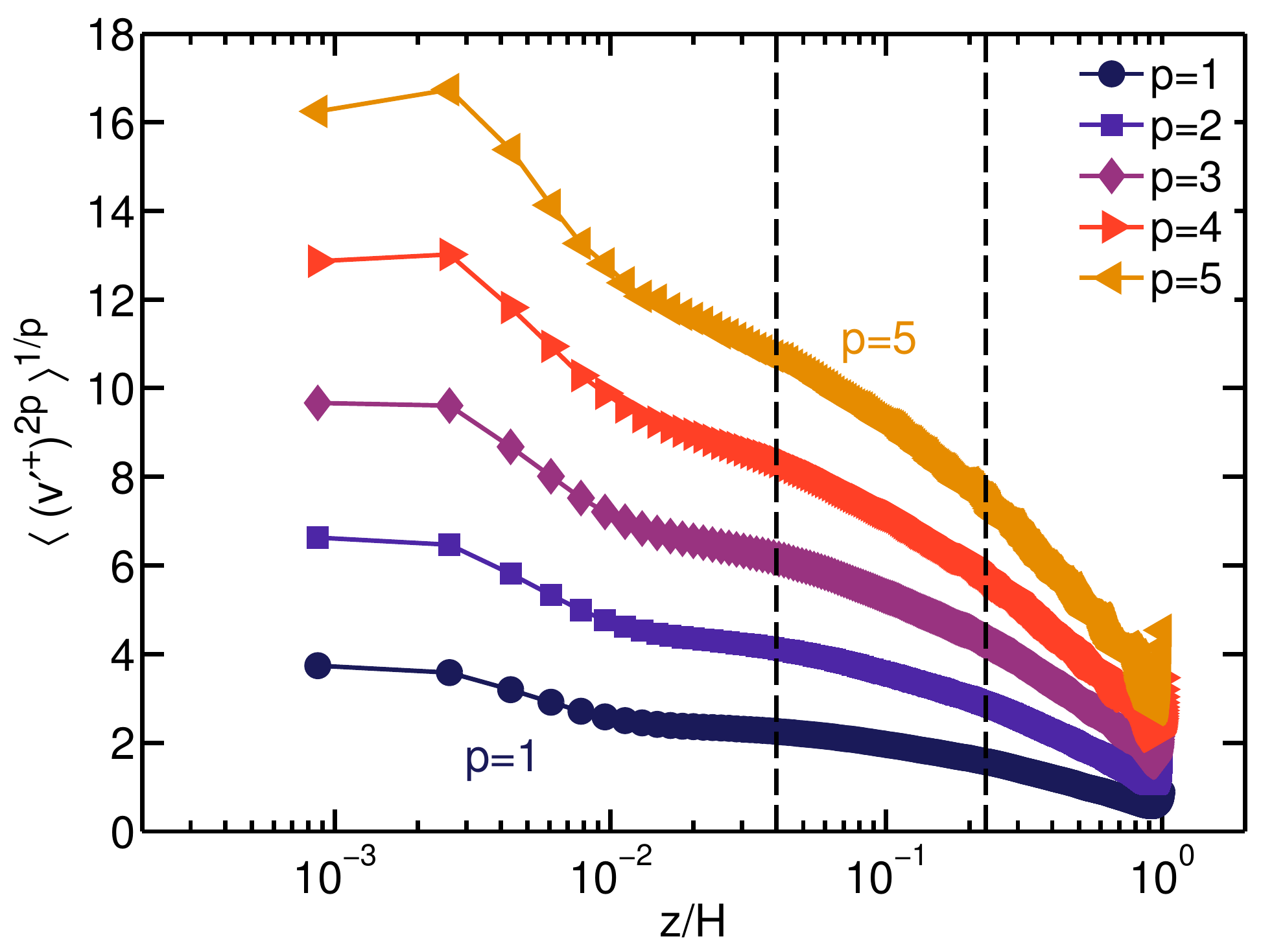}}
\subfigure[]{\includegraphics[width=0.49\textwidth]{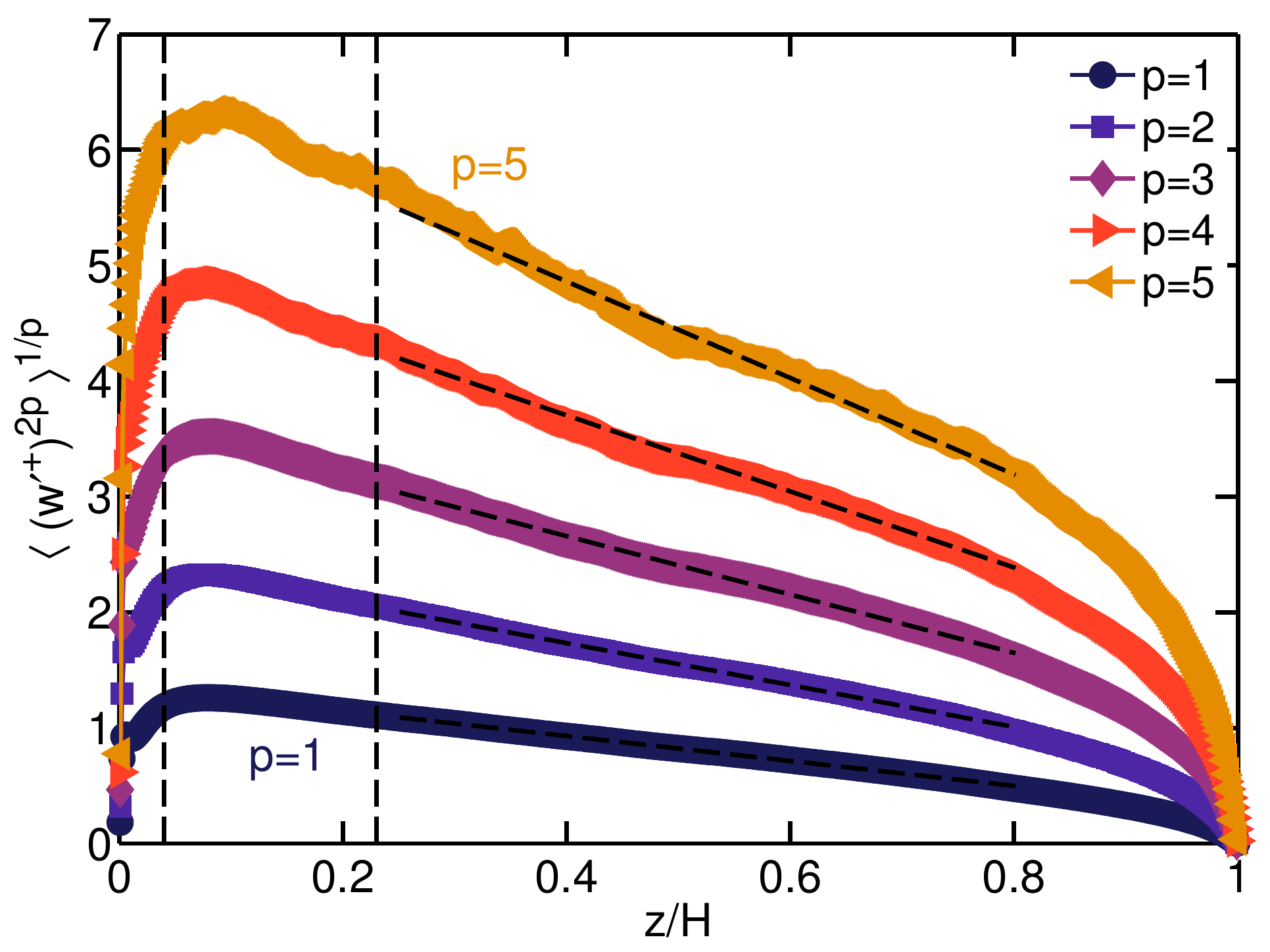}}
\caption{From bottom to top the figure indicates the moments of order $2p = 2$, $4$, $6$, $8$, and $10$ as a function of $z/H$ for the (a) spanwise and (b) vertical velocity fluctuations.}
\label{figure16}
\end{figure}

In contrast to the streamwise and spanwise velocity fluctuations, the higher-order moments for the vertical velocity fluctuations shown in figure \ref{figure15}b do not reveal any logarithmic region for the variance. Because the vertical velocity fluctuations seem to decrease linearly starting from the outer boundary of the logarithmic region up to approximately the top of the domain, these data are presented in a linear scale. In this region the higher-order moments of the vertical velocity seem to be fitted well by
\begin{equation} \label{equation6}
	\langle w^{\prime +2p} \rangle^{1/p}= - C_p \frac{z}{H} +D_p.
\end{equation}
We find that $C_p \approx 1.08, 1.81, 2.53, 3.29, 4.17$ for $p=1, 2, 3, 4, 5$, respectively. Interestingly, this means that $C_p$ increases almost linearly as function of $p$ (Gaussian prediction).

\section{Summary and conclusions} \label{section4}
We have used large-eddy simulations (LES) to study the scaling of higher-order moments in high Reynolds number turbulent wall-bounded flow. In the LES used here, the sub-grid scale stresses are modeled using a dynamic eddy viscosity sub-grid scale model, while the stress at the wall is modeled using a log-law based closure for rough surfaces. The focus of the study is not on comparing the performance of different sub-grid closures or to explore resolution requirements in detail. Instead, the focus is on exploring the capabilities of a more or less standard LES tool in predicting the generalized logarithmic laws that have been recently observed from data at very high Reynolds numbers. We also focus on the lower-limit (distance to the wall) of the generalized logarithmic law for moments observed near the wall in LES, on exploring whether trends observed in experimental data as function of Reynolds number (viscous scales) can be discerned when the latter are replaced by possible scalings with grid scale $\Delta/H$, SGS mixing length scale $C_s\Delta$, or roughness scale $z_0/H$. 

In terms of reproducing a logarithmic law for variances and higher-order moments of the streamwise velocity fluctuations, we find very good agreement between the LES and the experimental data, as long as a sufficiently fine resolution is used. In experiments the second and higher-order moments begin to deviate from the logarithmic law close to the wall due to viscous effects. Non-trivial dependencies on Reynolds numbers (i.e. viscous effects) were observed at significant distances from the wall (hundreds of wall units). In the LES, in which the viscous effects are not included explicitly, the higher-order moments also are found to deviate from the logarithmic law at some distance $z_b/H$ from the wall. Detailed tests show that for the LES this effect is coupled to the grid scale or (almost equivalently) to the SGS mixing length used in the simulation and that $z_b/H$ is (approximately) independent of $z_0/H$. While the simulations and comparisons with experimental data show that there might be a small dependence of $A_1$ on $z_0/H$, the observed trend is very weak compared to the uncertainties in the data and possible limitations of the simulations.

As all velocity components are available in the simulations, we also studied the spanwise and vertical velocity fluctuations. For the vertical velocity fluctuations we do not find any logarithmic regions. Instead, outside the logarithmic region of the streamwise velocity, the variance of vertical velocity fluctuations as well as appropriate roots of higher-order moments decrease approximately linearly with the distance from the wall. For the spanwise velocity fluctuations, the variance and the appropriate roots of higher-order moments do not show a very clear logarithmic region in the current dataset. However, we cannot exclude that significantly better resolved LES could reveal such a logarithmic region as the data for the spanwise velocity component are found to be more sensitive to numerical grid resolution than for the streamwise velocity component. The present analysis illustrates how the recently established logarithmic behavior of high-order moments in wall-bounded turbulence may be used to examine and test the accuracy of LES models with more rigor than only testing based on mean velocity profiles. It will be interesting to see how different sub-grid models, e.g.\  the standard Smagorinsky model, other eddy-viscosity models such as the \cite{vre97} or the WALE model \citep{nic99}, or the modulated gradient model \citep{lu10,lu13}, may perform in reproducing higher-order moments. We note that some models such as the modulated gradient model or the $k$-equation model can provide additional information about SGS variance. In addition, these tests can be used to asses how different types of wall model boundary conditions affect the results, see for example the trends shown in \cite{sto06} and the new insights about importance and coupling of stress fluctuations with outer-scale motions \citep{mar10}.

{\it Acknowledgment:} 
C.M. is grateful to I. Marusic for collaborations on wall-bounded turbulence and for making the Melbourne wind tunnel data available for comparisons. R.J.A.M.S. was supported by the `Fellowships for Young Energy Scientists' (YES!) of FOM, M.W. by DFG funding WI 3544/2-1, and C.M. by US National Science Foundation grants numbers CBET 1133800 and OISE 1243482. Most computations were performed with SURFsara resources, i.e.\ the Cartesius and Lisa clusters. This work was also supported by the use of the Extreme Science and Engineering Discovery Environment (XSEDE), which is supported by National Science Foundation grant number OCI-1053575.

\end{document}